\documentclass[10pt]{article}

\usepackage[utf8]{inputenc}
\usepackage[T1]{fontenc}
\usepackage{graphicx}
\usepackage{amsmath,amssymb}
\usepackage{booktabs}
\usepackage{hyperref}
\usepackage{natbib}
\usepackage[margin=1in]{geometry}
\usepackage{xcolor}
\usepackage{url}
\usepackage{enumitem}

\title{Attention Is Not Retention:\\
The Orthogonality Constraint in Infinite-Context Architectures}

\author{
Oliver Zahn$^1$, Matt Beton$^2$, and Simran Chana$^2$\\
\\
$^1$Independent Researcher \quad $^2$University of Cambridge
}

\date{}

\begin{document}

\maketitle

\begin{abstract}
The mammalian brain has developed a solution to a problem that current AI architectures are yet to solve: storing specific episodic facts without corrupting general semantic knowledge. Neuroscience explains this capacity through Complementary Learning Systems theory, which posits that biological memory requires two distinct subsystems, a fast hippocampal system for episodic storage using sparse, pattern-separated representations, and a slow neocortical system for extracting statistical regularities through distributed, overlapping representations. Current AI systems lack this architectural separation, attempting to serve both functions through neural weights alone.

We identify a fundamental geometric limit we term the \textit{Orthogonality Constraint}: reliable memory requires orthogonal keys, but semantic embeddings cannot be orthogonal because training clusters similar concepts together. The result is \textit{Semantic Interference} (connecting to what cognitive psychologists have long observed in human memory retrieval), where online neural memory systems writing facts into shared continuous parameters experience interference proportional to semantic density, collapsing to near-random accuracy within tens of semantically related facts. Though we use ``semantic'' to link to this literature, the interference mechanism operates in any embedding space where training clusters similar items. Through the lens of \textit{semantic density} ($\rho$), the mean pairwise cosine similarity of embeddings, we show that interference causes collapse at as few as $N=5$ facts at high density ($\rho > 0.6$) or $N \approx 20$--$75$ at moderate density. We validate this constraint across domains and modalities: 16,309 Wikipedia facts (real-world text), scientific measurements ($\rho = 0.96$, neural accuracy 0.02\% at $N=10{,}000$), and image embeddings ($\rho_{\text{intra}} = 0.82$, neural accuracy 0.05\% at $N=2{,}000$). This failure is geometric, not architectural; no regime exists in which increased model capacity, attention span, or training data can overcome interference when keys share semantic overlap.

The Orthogonality Constraint has immediate architectural consequences: reliable episodic memory requires discrete addressing. We propose \textit{Knowledge Objects} (KOs), structured facts with hash-based identity, controlled vocabularies, and explicit version chains, as the architecture that satisfies this constraint. On 16,309 Wikipedia facts, KO embedding retrieval achieves 45.7\% accuracy where Modern Hopfield Networks (the mathematical foundation of transformer attention) collapse to near-zero; hash-based retrieval maintains 100\% throughout. Production systems (Claude Memory, ChatGPT Memory) have partially adopted discrete storage but store \textit{unstructured text}, causing schema drift (40--70\% predicate consistency) and version ambiguity (0--100\% clean correction rates). Knowledge Objects address these limitations, providing the complete hippocampal component that current systems lack. Paired with neural weights for semantic reasoning, this bicameral architecture resolves the tension between fast episodic storage and slow statistical learning that pure neural approaches cannot escape.
\end{abstract}


\section{Introduction}

The brain separates two functions that AI architectures conflate: a hippocampus that rapidly encodes specific episodes with minimal interference, and a neocortex that slowly extracts statistical patterns across many experiences. This division is not an accident of biology but a solution to a fundamental computational problem, the same representations cannot simultaneously support fast, precise episodic storage and slow, distributed semantic learning \citep{mcclelland1995complementary}. Current AI architectures lack this separation, attempting to handle both episodic facts and semantic knowledge within the same neural substrate, whether context windows, parametric weights, or fast-weight associative memory. We argue that for online episodic storage via inner-product-addressed superposition in shared continuous parameters, overlap-driven interference imposes fundamental limits unless the system introduces explicit pattern separation.

\paragraph{Defining Memory.}
This paper distinguishes four forms of ``memory'' in AI systems that are often conflated in the literature. The first is \textit{context-window attention}, where facts remain visible within the current input sequence; this form is volatile (lost when the session ends) and incurs cost that grows with the amount of remembered material. The second is \textit{slow parametric memory} in model weights, where facts are compressed into parameters during pretraining or finetuning; this form offers vast capacity but is expensive to update and does not natively support surgical overwrite, deletion, or version semantics. The third is \textit{fast parametric memory} via online or associative mechanisms, where new facts are written at inference time into shared continuous parameters; this form enables instant writes but proves susceptible to interference under semantic density. The fourth is \textit{discrete addressable storage}, where facts are stored as explicit key-value records in external data structures; this form supports deterministic updates and avoids interference between entries entirely. Our analysis demonstrates that the first three forms cannot reliably support precise episodic memory under realistic conditions, and that the fourth, discrete addressing, is architecturally necessary for the hippocampal function.

\paragraph{Semantic Interference.}
We identify a core failure mode we term Semantic Interference: online neural memory collapses under semantic density. When facts share semantic overlap, for instance, ``Paris is the capital of France'' and ``Lyon is a city in France'', their representations interfere during retrieval because their embedding vectors occupy nearby regions of the continuous space. We formalize this phenomenon as the Orthogonality Constraint and demonstrate empirically that fast-weight and associative mechanisms collapse to near-random accuracy within tens of semantically related facts, far below theoretical capacity bounds that assume orthogonal keys. This limitation is not a bug to be fixed with scale or architectural refinement; it is a geometric inevitability arising from inner-product retrieval in shared continuous storage, where the training objective actively encourages the semantic clustering that causes storage interference.

\paragraph{Production Systems: Partial Solutions.}
Production systems have recognized this problem and moved toward discrete storage. ChatGPT Memory stores short timestamped entries like ``User is allergic to shellfish''\footnote{OpenAI Help Center, ``Memory FAQ,'' \url{https://help.openai.com/en/articles/8590148-memory-faq}} while Claude Memory stores memories in Markdown files.\footnote{Anthropic documentation describes a file-based architecture using CLAUDE.md files; see \url{https://docs.anthropic.com} and community analyses.} This represents progress, but only partial progress. These systems store \textit{unstructured text} extracted by LLMs rather than structured facts, and this causes two critical failure modes that we document empirically. First, \textit{schema drift}: predicate naming is inconsistent across extractions, the same fact might be stored as ``likes Python,'' ``prefers Python,'' or ``favorite language is Python'' depending on phrasing, breaking downstream lookup. Second, \textit{version ambiguity}: corrections fail to cleanly supersede prior values, with some models mentioning both old and new values rather than providing the canonical current state. Our experiments show 40--70\% schema consistency and 0--100\% clean correction rates across frontier models. Production systems have escaped the Orthogonality Constraint by adopting discrete storage, but they have not yet built the structured semantics required for reliable operation at scale.

\paragraph{The Solution: A Bicameral Architecture.}
We propose Knowledge Objects (KOs) as the missing hippocampal component in AI memory architecture. Knowledge Objects are discrete, typed memory units with controlled vocabularies that enforce schema consistency and explicit version chains that resolve ambiguity about current values. Paired with neural weights serving the neocortical role of slow semantic generalization, Knowledge Objects enable a true complementary learning architecture: fast, interference-free episodic storage operating alongside slow, distributed pattern extraction. A learned router classifies incoming queries and directs them to the appropriate subsystem based on whether they require precise factual retrieval or fuzzy semantic reasoning.

\paragraph{Contributions.}
This paper makes five contributions. First, we identify the \textit{Orthogonality Constraint}, a fundamental geometric limit on neural memory that causes collapse under semantic density, and validate it on real-world data across modalities (scientific measurements at $\rho = 0.96$, images at $\rho_{\text{intra}} = 0.82$). Second, we demonstrate that storage format determines retrieval accuracy at scale: structured subject-predicate-object facts achieve 45.7\% accuracy at $N=16{,}309$ facts while unstructured text achieves only 4.1\%, an 11$\times$ improvement with direct implications for production systems that currently store unstructured text. Third, we compare against Modern Hopfield Networks (the mathematical foundation of transformer attention), demonstrating that neural memory collapses catastrophically (97\%$\rightarrow$0\%) while discrete storage degrades gracefully (100\%$\rightarrow$46\%). Fourth, we diagnose schema drift (40--70\% consistency) and version ambiguity (0--100\% clean corrections) in current production systems, showing these are architectural limitations of unstructured storage. Fifth, we propose \textit{Knowledge Objects}, structured facts with hash-based identity, controlled vocabularies, and version chains, as the architecture that satisfies the Orthogonality Constraint, paired with a learned router achieving 97.8\% routing accuracy.

\paragraph{Scope and Non-Claims.}
The Orthogonality Constraint applies to any system that attempts online storage of discrete facts in shared continuous parameters. It does \textit{not} claim that LLMs cannot store facts; they obviously can, via slow pretraining. The constraint specifically concerns inference-time episodic memory. We do not claim that attention mechanisms are unnecessary, nor that larger models cannot memorize more facts during pretraining. Rather, we show that attention alone does not provide stable write-time memory under semantic interference; no regime exists in which increased attention capacity alone improves long-term factual retention under semantic density. Similarly, we do not claim that RAG is useless; we show that naive similarity-only RAG lacks the identity, schema, and version semantics required for true episodic memory, and that adding these capabilities effectively introduces KO-like structure.

\paragraph{Paper Structure.}
Section~\ref{sec:related} reviews related work. Section~\ref{sec:theory} presents the theoretical framework. Section~\ref{sec:methods} describes Knowledge Objects and the hybrid architecture. Section~\ref{sec:experiments} presents experimental validation. Section~\ref{sec:discussion} discusses implications.

\begin{figure}[t]
    \centering
    \includegraphics[width=\textwidth]{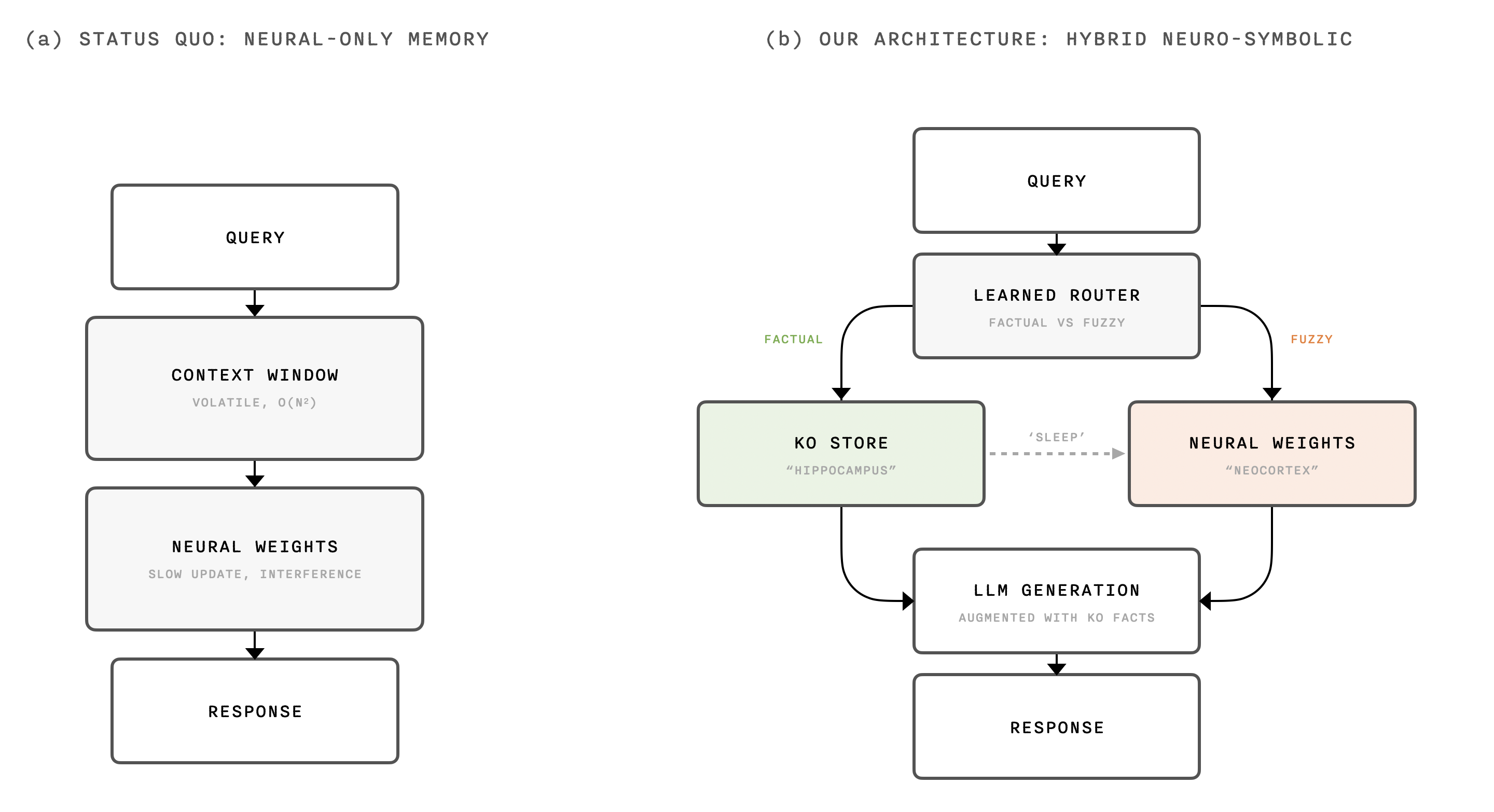}
    \caption{\textbf{The Memory Architecture Gap.} \textit{(a)} Pure neural approaches to online memory, context windows and fast weights, face the Orthogonality Constraint: interference grows with semantic density, causing collapse. Production systems have adopted discrete external storage but use unstructured text, causing schema drift and version ambiguity. \textit{(b)} Knowledge Objects provide structured discrete storage with hash-based identity, controlled vocabularies, and version chains. This is the architecture that satisfies the Orthogonality Constraint while providing the semantic guarantees that production systems lack.}
    \label{fig:architecture_comparison}
\end{figure}

\section{Related Work}
\label{sec:related}

\paragraph{Associative Memory Capacity.}
Classical work established fundamental limits on neural memory. Hopfield networks store approximately $0.14N$ patterns before ``blackout catastrophe'' \citep{hopfield1982neural, mceliece1987capacity}. Ramsauer et al. \citep{ramsauer2021hopfield} proved that transformer attention is equivalent to a modern Hopfield network update, inheriting these capacity constraints. Schlag et al. \citep{schlag2021linear} showed linear transformers enter an ``overcapacity regime when sequence length exceeds key dimension,'' establishing $O(d)$ bounds. Our work extends this analysis by introducing \textit{semantic density} as a failure predictor: we show collapse occurs far below theoretical capacity when stored patterns share embedding similarity.

\paragraph{Episodic Memory for LLM Agents.}
Recent work has argued that LLM agents require separate episodic memory modules to bridge parametric and context memory \citep{schmidgall2025episodic}. While we share the conclusion that episodic memory separation is essential, our contribution is distinct: we provide mathematical and empirical evidence that such separation is necessary due to geometric interference limits (the Orthogonality Constraint), not merely beneficial for agent capabilities. The growing consensus around episodic memory modules validates our thesis; we supply the formal foundation explaining why this architecture emerges.

\paragraph{Semantic Interference in Human and Machine Memory.}
We term our core phenomenon \textit{Semantic Interference}, connecting to what cognitive psychologists have long observed in human memory retrieval: semantically similar items interfere with each other during recall \citep{loewenstein2004semantic, muller1900experimentelle}. M{\"u}ller and Pilzecker identified this effect in 1900; over a century of subsequent research has confirmed that retrieval accuracy degrades when target items share semantic features with competitors. Our contribution is providing the mathematical framework explaining \textit{why} this occurs in any associative memory system: the orthogonality constraint (Eq.~2) is violated when semantically similar items cluster in embedding space.

\paragraph{Neural Episodic Memory.}
Recent architectures like Titans \citep{munkhdalai2025titans} and Linear Transformers \citep{katharopoulos2020transformers} learn ``fast weights'' at test time, treating memory as gradient optimization. While Titans reports 98.8\% accuracy on associative recall, it provides no capacity bounds and does not characterize failure modes under semantic density. We show that the core mechanism underlying such architectures, superposition of key-value pairs in shared continuous parameters, leads to collapse when storing semantically related facts.

\paragraph{Nested Learning and Associative Memory.}
Behrouz et al. \citep{behrouz2025nested} formalize the view that neural architectures and optimizers are nested systems of associative memories, each compressing its own ``context flow'' at different update frequencies. Their framework elegantly unifies learning dynamics across multiple timescales, drawing on the same neuroscientific inspiration (brain oscillations, multi-timescale processing) that motivates our work. However, their analysis assumes that effective memory can be achieved through learned compression in continuous parameters. Our Orthogonality Constraint identifies a fundamental limit they do not address: when stored keys share semantic similarity, interference during retrieval becomes geometrically inevitable regardless of how the memory is trained. Their ``Continuum Memory System'' proposes a spectrum of update frequencies; our contribution is explaining why at least one level of this spectrum must employ discrete addressing to achieve interference-free episodic storage.

\paragraph{Memory-Augmented Neural Networks.}
Neural Turing Machines \citep{graves2014neural} and Differentiable Neural Computers introduced external memory banks with soft attention-based addressing. Memory Networks \citep{weston2014memory} demonstrated multi-hop reasoning over explicit memory slots. These systems use continuous addressing (soft attention over memory locations), which we show suffers from the same interference problem as in-weight storage when keys are semantically similar. Our Knowledge Objects use discrete, hash-based addressing that guarantees zero interference.

\paragraph{Retrieval Augmented Generation.}
RAG systems \citep{lewis2020retrieval} offload memory to external stores, retrieving from corpora at inference time. While modern vector databases support near-real-time upserts, RAG's architecture is fundamentally \textit{append-first}: new facts coexist with old ones, and there is no native mechanism to identify a canonical ``current value'' or cleanly supersede outdated information. Knowledge Objects differ in providing \textit{key-based identity} for deterministic overwrite, \textit{controlled vocabularies} for schema consistency, and \textit{version chains} for explicit supersession semantics.

\paragraph{Graph-Based Knowledge Systems.}
GraphRAG \citep{edge2024graphrag} constructs knowledge graphs from documents, enabling structured traversal. Our Knowledge Objects share the insight that typed, structured memory outperforms unstructured chunk retrieval. We differ in three ways: (1) key-based identity enabling deterministic overwrite rather than append-only indexing, (2) learned routing to classify query intent, and (3) version chains enabling temporal queries and clean supersession of outdated facts.

\paragraph{Complementary Learning Systems.}
Our architecture is directly inspired by Complementary Learning Systems (CLS) theory from neuroscience \citep{mcclelland1995complementary, kumaran2016learning}. CLS proposes that biological memory requires two systems with fundamentally different properties: a hippocampal system characterized by fast learning, sparse and discrete representations, and pattern separation that enables storage of specific episodes without interference; and a neocortical system characterized by slow learning, distributed and overlapping representations, and pattern completion that extracts statistical regularities through gradual exposure across many experiences. The hippocampus solves exactly the problem we address: storing arbitrary new facts quickly without corrupting existing memories.

Critically, the hippocampus achieves this through a specific mechanism: the \textit{Dentate Gyrus} performs \textit{pattern separation}, taking overlapping sensory inputs and mapping them to sparse, orthogonal neural representations. This forces semantically similar inputs (``kitchen today'' vs.\ ``kitchen yesterday'') onto distinct neural populations, preventing interference. Our hash function $\text{hash}(\text{subject} \| \text{predicate})$ serves exactly this role: it takes semantically dense embedding vectors (which overlap in continuous space) and maps them to discrete addresses (which are orthogonal by construction). Knowledge Objects thus implement an \textit{artificial Dentate Gyrus}, a pattern separation layer that overcomes the Orthogonality Constraint by forcing storage into non-interfering addresses. Neural weights play the neocortical role of slow semantic generalization, and sleep consolidation (which transfers hippocampal memories to cortex) maps to our proposed mechanism for promoting stable KO facts into model weights.

\section{Theoretical Framework: The Orthogonality Constraint}
\label{sec:theory}

Why does online neural memory collapse under semantic density? We analyze the interference dynamics of a stylized Linear Associative Memory (LAM) module, as a proxy for test-time ``fast-weight'' and associative memory mechanisms (e.g., Titans-style updates) that write new facts by superposition into a shared continuous matrix at inference time. Throughout this section, the Orthogonality Constraint refers to separation in the key/query embedding space used to address memory (dimension $d$, typically the hidden size), \textit{not} to the total number of parameters of a pretrained LLM, nor to the ability of slow parametric weights to encode broad world knowledge after large-scale training.

Given $N$ stored fact pairs $\{(\mathbf{k}_i, \mathbf{v}_i)\}_{i=1}^N$, the memory matrix is:
\begin{equation}
    \mathbf{M} = \sum_{i=1}^{N} \mathbf{v}_i \otimes \mathbf{k}_i^T
\end{equation}

When retrieving fact $j$ by querying with $\mathbf{k}_j$:
\begin{equation}
    \mathbf{M} \cdot \mathbf{k}_j = \underbrace{\mathbf{v}_j (\mathbf{k}_j \cdot \mathbf{k}_j)}_{\text{signal}} + \underbrace{\sum_{i \neq j} \mathbf{v}_i (\mathbf{k}_i \cdot \mathbf{k}_j)}_{\text{interference}}
\end{equation}

The retrieval output is the sum of the intended signal ($\mathbf{v}_j$) plus an interference term from every other stored fact. Eq.~(2) shows retrieval is the sum of a target \textit{signal} and a multi-item \textit{interference} term. Increasing $N$ under non-negligible semantic density $\rho$ causes interference to dominate, so the failure mode is \textit{stability/isolation under superposition} (cross-talk), not a claim that pretrained weights cannot store large amounts of knowledge. If keys were orthogonal ($\mathbf{k}_i \cdot \mathbf{k}_j = 0$ for $i \neq j$), interference would vanish. But real semantic embeddings have non-zero semantic density:
\begin{equation}
    \rho = \frac{1}{N(N-1)} \sum_{i \neq j} \cos(\mathbf{k}_i, \mathbf{k}_j)
\end{equation}

Retrieval fails when interference overwhelms signal. With $N$ facts at density $\rho$, the expected interference magnitude scales as $O(N \cdot \rho)$. This predicts: (1) collapse accelerates with $N$, and (2) higher $\rho$ causes earlier collapse. We note that this scaling argument is approximate: the tightest theoretical bound depends on both the mean and variance of pairwise cosines, and may scale as $O(\sqrt{N} \cdot \sigma_\rho)$ under certain distributional assumptions. However, our measure $\rho$ (mean pairwise cosine) serves as an effective empirical predictor, and our experiments (Figure~\ref{fig:density_collapse}) validate the predictive relationship between $\rho$ and collapse threshold $N_{50}$ regardless of the precise theoretical bound.

\paragraph{Information-Theoretic Interpretation.}
Eq.~(2) admits a signal-to-noise interpretation: the target fact contributes signal $\mathbf{v}_j(\mathbf{k}_j \cdot \mathbf{k}_j)$, while all other facts contribute interference noise $\sum_{i \neq j} \mathbf{v}_i(\mathbf{k}_i \cdot \mathbf{k}_j)$. The retrieval signal-to-noise ratio (SNR) thus scales inversely with $N \cdot \rho^2$. This connects Semantic Interference to Shannon-Hartley channel capacity: just as a communication channel fails when noise power exceeds signal power, associative memory fails when interference overwhelms the target. The critical difference from classical Hopfield capacity bounds is that we measure \textit{semantic} density rather than assuming random patterns: semantic embeddings, by design, exhibit far higher density than random vectors would.

\paragraph{Why High-Dimensional Capacity Bounds Don't Help.}
A natural objection is that high-dimensional spaces can accommodate exponentially many near-orthogonal vectors. Johnson-Lindenstrauss-style bounds suggest that for embedding dimension $d$ and maximum pairwise similarity $\varepsilon$, one can fit:
\begin{equation}
    N \lesssim \exp\left(\frac{\varepsilon^2 \cdot d}{4}\right)
\end{equation}
For $d=7000$ (DeepSeek-scale) and $\varepsilon=0.1$, this permits $\sim 4 \times 10^7$ vectors. Why then does collapse occur at $N \approx 20$?

The problem is that semantic embeddings cannot exploit this geometric capacity. The training objective for embedding models \textit{requires} similar concepts to cluster together, this is what enables generalization and transfer. ``Paris is the capital of France'' and ``Lyon is a city in France'' \textit{must} be nearby in embedding space because they share words, appear in similar contexts, and predict similar continuations.

The Orthogonality Constraint thus arises not from geometric limits on $d$, but from the fundamental tension between two objectives that cannot be simultaneously satisfied. The \textit{compression objective}, what embeddings are trained for, requires grouping semantically similar content together to enable generalization and transfer learning. The \textit{storage objective}, what factual recall requires, demands separating distinct facts to enable interference-free retrieval. These objectives are fundamentally opposed: the same geometric property (clustering similar concepts) that makes embeddings useful for prediction makes them unreliable for storage.

\paragraph{But Doesn't Pretraining Solve This?}
One might argue that pretraining already solves this problem: the model learns to ``squeeze all possible semantic meanings into $d$-dimensional space'' efficiently. If pretraining can pack exponentially many concepts, why can't it pack facts?

The answer is that pretraining optimizes for \textit{prediction}, not \textit{storage}. These objectives are fundamentally opposed:

\begin{center}
\begin{tabular}{lcc}
\toprule
& \textbf{Prediction} & \textbf{Storage} \\
\midrule
Goal & Predict next token & Retrieve exact fact \\
Similar inputs & Should produce similar outputs & Must remain distinguishable \\
``France facts'' & Should cluster (enables transfer) & Must be separable (enables lookup) \\
Trained for & Generalization & Precision \\
\bottomrule
\end{tabular}
\end{center}

The embedding model is \textit{trained} to put ``Paris is capital of France'' near ``Lyon is city in France'' because they share words, appear in similar contexts, and predict similar continuations. This clustering is a \textit{feature} for generalization, it enables the model to transfer knowledge about France to new contexts. But it is a \textit{bug} for factual lookup, it causes ``Paris'' and ``Lyon'' to interfere with each other during retrieval.

If pretraining had successfully orthogonalized facts, we would not observe: (1) the ``Lost in the Middle'' phenomenon, (2) the interference collapse we document, (3) models hallucinating similar-but-wrong facts, or (4) semantic drift in long-context agents. But we observe all of these. The JL capacity exists in theory, but the training objective does not exploit it for fact storage, it exploits it for pattern compression.

Real-world knowledge exhibits $\rho \approx 0.3$--$0.7$, not $\rho \approx 0$. This is why collapse occurs at $N \approx 20$, far below both the JL bound and the classical Hopfield capacity of $\sim 0.14d \approx 54$ for $d=384$.

\paragraph{Discrete Memory: True Orthogonality.}
In discrete memory, keys are hash addresses; if $\mathbf{k} \neq \mathbf{k}_{\text{old}}$, overlap is exactly 0 by construction. Semantic Interference (the accuracy gap between discrete and neural memory) is thus a direct consequence of accumulated interference in continuous representations.

\begin{figure}[t]
    \centering
    \includegraphics[width=\textwidth]{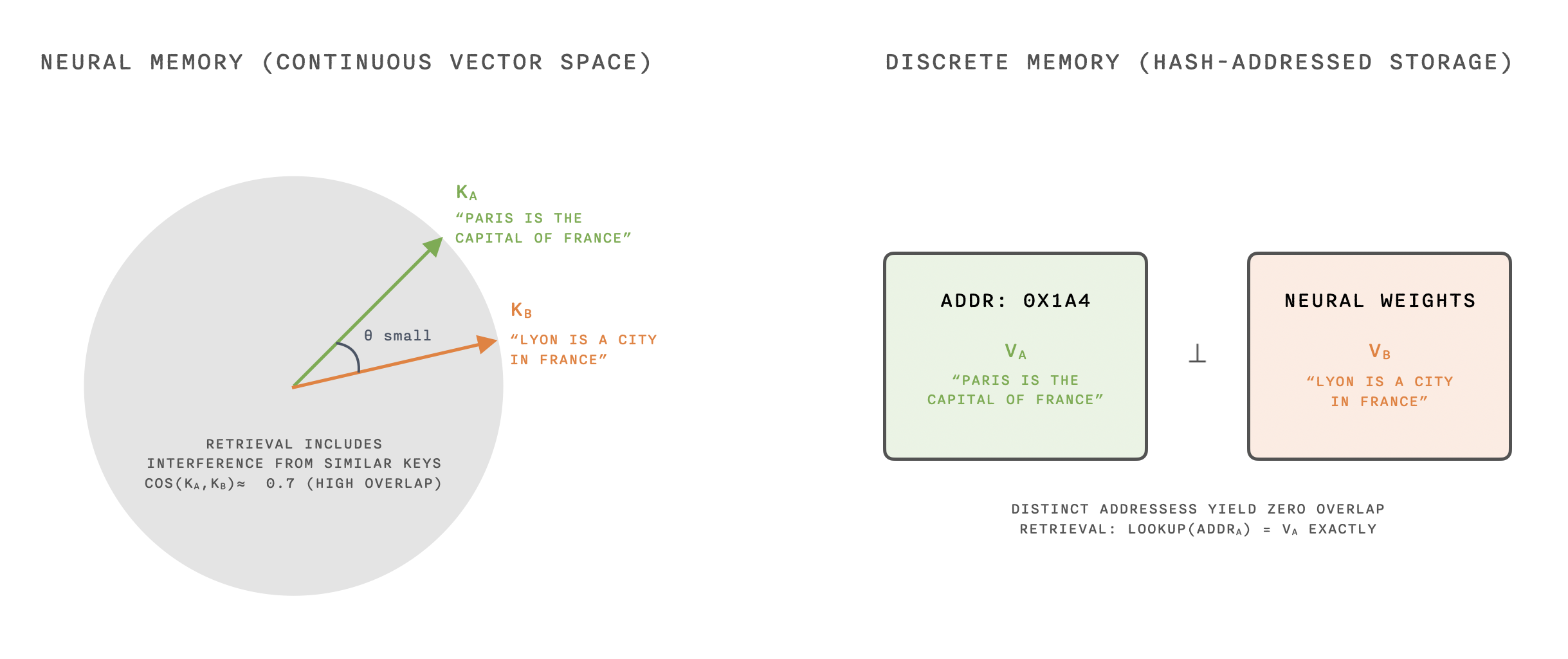}
    \caption{\textbf{The Orthogonality Constraint.} \textit{Left:} In neural memory, semantically related facts (here, two statements about French cities) produce embedding vectors with small angular separation ($\cos\theta \approx 0.7$). When retrieving one fact, the system also partially retrieves others with similar keys, interference that grows with corpus size $N$ and semantic density $\rho$, causing collapse by $N \approx 20$--$75$ depending on density. \textit{Right:} Discrete memory assigns unique hash addresses to each fact; overlap between addresses is exactly zero by construction. Retrieval returns $\mathbf{v}_A$ with no interference regardless of how many other facts share semantic similarity.}
    \label{fig:interference}
\end{figure}

\section{Methodology}
\label{sec:methods}

This section presents our methodology in detail, beginning with the neural baseline that serves as our theoretical comparator, followed by the Knowledge Object architecture that constitutes our primary contribution, and concluding with the hybrid system that combines both approaches through learned routing.

\subsection{Neural Memory Baseline: Linear Associative Memory}

To represent gradient-based episodic memories such as Titans and Linear Transformers, we implement a Test-Time Training (TTT) Linear Memory, a simplified proxy that captures the essential dynamics of systems that learn via gradient updates to shared continuous weights. The storage mechanism employs a weight matrix $\mathbf{M} \in \mathbb{R}^{d \times d}$ to store all facts; given a key-value pair $(\mathbf{k}, \mathbf{v})$, the matrix updates according to
\begin{equation}
    \mathbf{M}_{t+1} = \lambda \mathbf{M}_t + \eta (\mathbf{v} \otimes \mathbf{k}^T)
\end{equation}
where $\lambda = 0.99$ is a decay factor and $\eta = 0.1$ is the learning rate. Retrieval proceeds by computing $\hat{\mathbf{v}} = \mathbf{M} \mathbf{q}$ for a query $\mathbf{q}$ and decoding via nearest-neighbor matching against stored values. This baseline captures the core failure mode we analyze: gradient updates to shared weights cause interference when keys overlap in the embedding space. Production systems may incorporate additional mechanisms such as normalization and regularization, but the fundamental constraint, that semantically similar keys produce blended retrievals, remains inherent to any architecture that stores facts via superposition.

\subsection{Knowledge Objects: Online Episodic Memory}
\label{sec:ko}

Knowledge Objects constitute the core methodological contribution of this work. Unlike RAG, which retrieves from static pre-indexed corpora, KOs provide \textit{online episodic memory}: facts written during inference become available immediately, without requiring batch re-indexing. This capability addresses a fundamental limitation of RAG systems that has received insufficient attention in the literature.

\subsubsection{The RAG Limitation}

Consider a conversational agent in which a user states ``My name is Alice.'' Modern vector databases can perform near-real-time upserts, so the embedding could be added quickly. However, RAG's architectural limitations run deeper than indexing latency. First, RAG is \textit{append-first}: adding the new fact does not remove or supersede any prior statement about the user's name, so both old and new values may be retrieved. Second, RAG lacks \textit{key-based identity}: there is no canonical address for ``the user's name'' that can be surgically updated; facts are identified only by embedding similarity, which is fuzzy. Third, RAG provides no \textit{schema guarantees}: the fact might be stored as ``user's name is Alice'' or ``Alice is the user's name'' depending on phrasing, breaking downstream systems that expect consistent structure. Fourth, RAG has no native \textit{version semantics}: if the user later says ``Actually, call me Alicia,'' the system has no mechanism to mark the prior value as superseded, both will be retrieved with similar scores.

These limitations are not addressed by standard similarity-only RAG; resolving them requires adding explicit identity, schema, and version semantics, effectively introducing KO-like structure. Knowledge Objects provide these capabilities natively: hash-based addressing provides key-based identity for surgical updates, controlled vocabularies enforce schema consistency, version chains provide explicit supersession semantics, and the combination enables true online episodic memory where facts are available immediately with deterministic overwrite behavior.

\subsubsection{Knowledge Object Data Structure}

Each Knowledge Object is represented as a typed tuple of the form
\begin{equation}
    \text{KO} = (\text{id}, \text{subject}, \text{predicate}, \text{object}, \text{embedding}, \text{provenance})
\end{equation}
where the \textit{id} is a unique identifier computed as $\text{hash}(\text{subject} \| \text{predicate})$, the \textit{subject} denotes the entity about which the fact pertains (such as ``user'', ``patient\_MR'', or ``France''), the \textit{predicate} specifies the attribute or relationship (such as ``name'', ``capital'', or ``tumor\_size''), and the \textit{object} contains the value itself (such as ``Alice'', ``Paris'', or ``3cm''). The \textit{embedding} is a dense vector computed from the concatenation of subject, predicate, and object, enabling semantic search, while the \textit{provenance} field captures metadata including source, timestamp, confidence score, and version pointer.

The Knowledge Object architecture supports richer content structures including multi-sentence claims, hypotheses, and reasoning traces; we use subject-predicate-object facts here to isolate the retrieval mechanism from content complexity.

\subsubsection{Storage: Pattern Separation via Hash Addressing}

Knowledge Objects are stored in a hash map keyed by $\text{hash}(\text{subject} \| \text{predicate})$. This hash function implements \textit{pattern separation}: it takes semantically overlapping inputs (embedding vectors with $\cos\theta > 0$) and maps them to discrete addresses (which are orthogonal by construction, with overlap exactly 0). The resulting storage provides constant-time insertion and update, constant-time exact lookup when the subject and predicate are known, and zero interference between facts since distinct keys map to distinct addresses and updating one fact cannot corrupt another. This last property matters most: whereas neural memory systems store all facts in shared continuous parameters where updates necessarily interact, hash-based storage guarantees complete isolation between memory entries, the same guarantee the Dentate Gyrus provides for biological episodic memory.

\subsubsection{Retrieval: Semantic Search over Embeddings}

For queries that do not specify an exact subject-predicate pair, we perform $k$-nearest neighbor search over the embedding space. The process begins by encoding the query as $\mathbf{q} = \text{Encoder}(\text{query})$, then finding the top-$k$ Knowledge Objects by cosine similarity via $\text{argmax}_k \cos(\mathbf{q}, \mathbf{e}_i)$, and finally returning the matched KOs to augment the LLM's response. This hybrid approach, exact lookup when possible, semantic search otherwise, combines the precision of structured databases with the flexibility of neural retrieval.

\subsubsection{Two Retrieval Modes: A Critical Distinction}

Knowledge Objects support two fundamentally different retrieval modes with distinct accuracy and latency characteristics, and conflating them leads to confusion about system capabilities.

\textit{Hash-based retrieval} applies when the subject and predicate are known (e.g., a structured query for \texttt{(France, capital, ?)}). For predicates that are \textit{functional}, where each subject has exactly one value, hash lookup achieves 100\% accuracy with $O(1)$ latency. Multi-valued predicates (a country has multiple neighbors, a person has multiple employers over time) require standard extensions such as list storage or composite keys.

\textit{Embedding-based retrieval} applies when queries are natural language (e.g., ``What is the capital of France?''). This mode relies on the embedding model's ability to map the query near the correct fact in vector space. Accuracy depends on embedding quality and semantic density of the corpus, we observe approximately 46\% top-1 accuracy on Wikipedia facts, with 78\% top-5 recall. Latency scales as $O(N)$ for brute-force search or $O(\log N)$ with approximate nearest neighbor indices such as HNSW.

The experiments in Section~\ref{sec:experiments} primarily evaluate embedding-based retrieval because it represents the harder case: users issue natural language queries, and the system must locate relevant facts without knowing the exact keys. Production deployments can improve effective accuracy through query parsing (converting natural language to structured queries when possible), multi-value hash storage, reranking (using a cross-encoder to rescore top-$k$ candidates), or hybrid routing (attempting hash lookup first, falling back to embedding search).

\subsubsection{The KO Compiler: Extracting Structured Facts}

Raw text must be compiled into structured Knowledge Objects, a process we accomplish through an LLM-based compiler. The compiler receives natural language input and extracts factual statements as subject-predicate-object facts, using snake\_case for predicates to ensure consistency and including confidence scores for each extraction. For instance, given the input ``Dr. Sarah Chen is a 42-year-old cardiologist at MGH. Her phone is 617-555-0123,'' the compiler produces four facts: \texttt{(Dr\_Sarah\_Chen, age, 42)}, \texttt{(Dr\_Sarah\_Chen, occupation, cardiologist)}, \texttt{(Dr\_Sarah\_Chen, employer, MGH)}, and \texttt{(Dr\_Sarah\_Chen, phone\_number, 617-555-0123)}.

This extraction step is where \textit{schema drift} can occur: without controlled vocabularies constraining predicate choices, an LLM might extract the same fact as \texttt{phone\_number}, \texttt{phone}, or \texttt{contact\_phone} depending on phrasing. Experiment 7 quantifies this drift across frontier models. Our retrieval experiments (Experiments 1--6, 8) use pre-structured data (Wikipedia facts, templated scientific measurements) to isolate the retrieval mechanism from extraction variability, this represents the best-case scenario where extraction is perfect, establishing an upper bound on system performance.

\subsubsection{Surgical Updates: $O(1)$ Correction}

When a fact changes, KO performs a surgical update by computing the identifier $\text{id} = \text{hash}(\text{subject} \| \text{predicate})$, optionally archiving the old value with a version pointer for history preservation, and writing the new value at the same address. This operation completes in $O(1)$ time regardless of corpus size. By contrast, RAG systems must append the new fact to a pending queue, re-index the entire corpus or affected partition, and handle the conflict period during which old and new facts coexist until re-indexing completes, a process that scales with corpus size and introduces temporal inconsistency.

\subsubsection{Version Chains: Temporal Queries}

Knowledge Objects optionally preserve version history, maintaining a linked list of prior values with timestamps. For example, a user's favorite food might have evolved through versions: spaghetti (January 2024, archived), sushi (June 2024, archived), and pizza (January 2025, current). This structure enables temporal queries: asking ``What is my favorite food?'' returns the current value (pizza), asking ``What was my favorite food before?'' returns the previous version (sushi), and asking ``What changed since January?'' triggers traversal of the version chain. Neural memory cannot readily support such queries because weight updates destroy prior states irrecoverably, there is no ``undo'' operation for gradient descent.

\subsubsection{Key Properties Summary}

Table~\ref{tab:ko-rag} summarizes the architectural differences between Knowledge Objects and conventional RAG systems.

\begin{table}[h]
\centering
\caption{Knowledge Objects vs RAG: Architectural comparison.}
\label{tab:ko-rag}
\begin{tabular}{lcc}
\toprule
\textbf{Property} & \textbf{RAG} & \textit{Knowledge Objects} \\
\midrule
Fact identity & Embedding similarity (fuzzy) & Hash key (deterministic) \\
Update semantics & Append (old persists) & Supersede (clean overwrite) \\
Schema control & None (free-form chunks) & Controlled vocabulary \\
Data structure & Unstructured chunks & Typed (subject, predicate, object) \\
Version handling & Both values retrieved & Explicit version chains \\
Temporal queries & Not supported & Traverse version history \\
\bottomrule
\end{tabular}
\end{table}

\subsubsection{Illustrative Example: A Medical Research Assistant}

To make the architecture concrete, consider a medical research assistant that helps a physician track patient cases over months of interaction. In a first session during January, the physician states: ``Patient M.R., 67-year-old male, presented with persistent cough and 15-pound weight loss. CT scan showed 3cm mass in right upper lobe. Biopsy confirmed non-small cell lung cancer, stage IIB.'' The KO compiler extracts structured facts, \texttt{(patient\_MR, age, 67)}, \texttt{(patient\_MR, diagnosis, NSCLC\_IIB)}, \texttt{(patient\_MR, tumor\_size, 3cm)}, and \texttt{(patient\_MR, tumor\_location, right\_upper\_lobe)}, which are written immediately to the KO store and become available for retrieval in subsequent queries without waiting for batch reindexing.

In a second session during March, the physician asks: ``What was M.R.'s tumor size at diagnosis?'' The router classifies this as a factual query, and KO retrieval returns \texttt{3cm} deterministically via hash lookup on \texttt{(patient\_MR, tumor\_size)}. The physician then reports: ``Post-chemo scan shows tumor reduced to 1.2cm.'' The system performs a surgical update in which the new value supersedes the prior one, but the version chain preserves history, enabling later queries such as ``How has M.R.'s tumor responded to treatment?'' that require temporal reasoning over the fact's evolution.

\textit{Session 15 (November):} The physician asks: ``Summarize M.R.'s treatment trajectory.'' The system retrieves the full version chain for \texttt{patient\_MR}, constructs a temporal narrative, and generates: ``M.R. presented in January with 3cm NSCLC in the right upper lobe. Following chemotherapy, tumor reduced to 1.2cm by March...''

This scenario illustrates the capabilities that distinguish Knowledge Objects from alternatives: facts become available immediately without waiting for reindexing (zero-latency writes), tumor size changes can be recorded without corrupting diagnosis or age information (surgical updates), queries about historical states can traverse the version chain (temporal queries), and factual queries hit the KO store while requests like ``explain the treatment options'' pass to the LLM (hybrid routing). A pure neural system would suffer interference as hundreds of patient facts accumulate, while a pure RAG system could not update tumor size without re-indexing the corpus.

\subsection{Hybrid System: The Learned Router}

Neither pure discrete memory nor pure neural generation handles all queries effectively. Factual queries such as ``What is the capital of France?'' benefit from KO's interference-free retrieval, while fuzzy queries such as ``Write a poem about Paris'' should pass directly to the LLM. The hybrid architecture addresses this through a learned routing mechanism.

\subsubsection{Architecture Overview}

The system operates in four stages: first, the query is embedded using a sentence transformer; second, a learned router classifies the query as either ``Factual'' or ``Fuzzy''; third, factual queries follow the fast path, retrieving from the KO store and augmenting the LLM context with the results; and fourth, fuzzy queries follow the slow path, passing directly to the LLM without KO retrieval.

\subsubsection{Why Not Static Thresholds?}

Initial experiments employed a static confidence threshold ($\tau=0.75$), where the system would use a KO match if its cosine similarity exceeded the threshold and fall back to the LLM otherwise. This approach failed under load because semantic crowding reduced cosine similarities below the threshold even for correct matches, triggering inappropriate fallback to the LLM when KO retrieval would have been appropriate.

\subsubsection{The Learned Router}

We experimented with two router variants: (1) logistic regression on raw query embeddings (384-dimensional vectors from the sentence transformer), and (2) logistic regression on handcrafted surface features extracted from the query text. Both approaches use a training set of 30 examples comprising 15 factual queries (such as ``What is X?'' and ``Who created Y?'') and 15 fuzzy queries (such as ``Write a poem'' and ``Explain the concept''). We include the router to demonstrate feasibility of intent separation; we do not claim this specific classifier is optimal, and production deployments would likely benefit from larger training sets and more sophisticated classifiers. The surface-feature variant, which we report in our experiments, extracts features such as: presence of interrogative words (what, who, when, where), question mark termination, proper noun indicators, word count, and creative/explanatory verbs (write, help, suggest, explain). This variant achieved slightly better accuracy and interpretability, though both approaches exceeded 95\% accuracy. We use logistic regression as the simplest router; in Experiment 5 (Table~\ref{tab:mixed}) we also evaluate other lightweight classifiers (Random Forest, SVM, Gradient Boosting) on the same feature set to show that the routing task is robustly learnable. What makes this work is that the router examines \textit{linguistic structure} rather than embedding similarity scores, making it immune to vector space crowding: a query like ``What is the capital of France?'' is classified as factual regardless of how crowded the embedding space becomes.

\begin{figure}[t]
    \centering
    \includegraphics[width=0.9\textwidth]{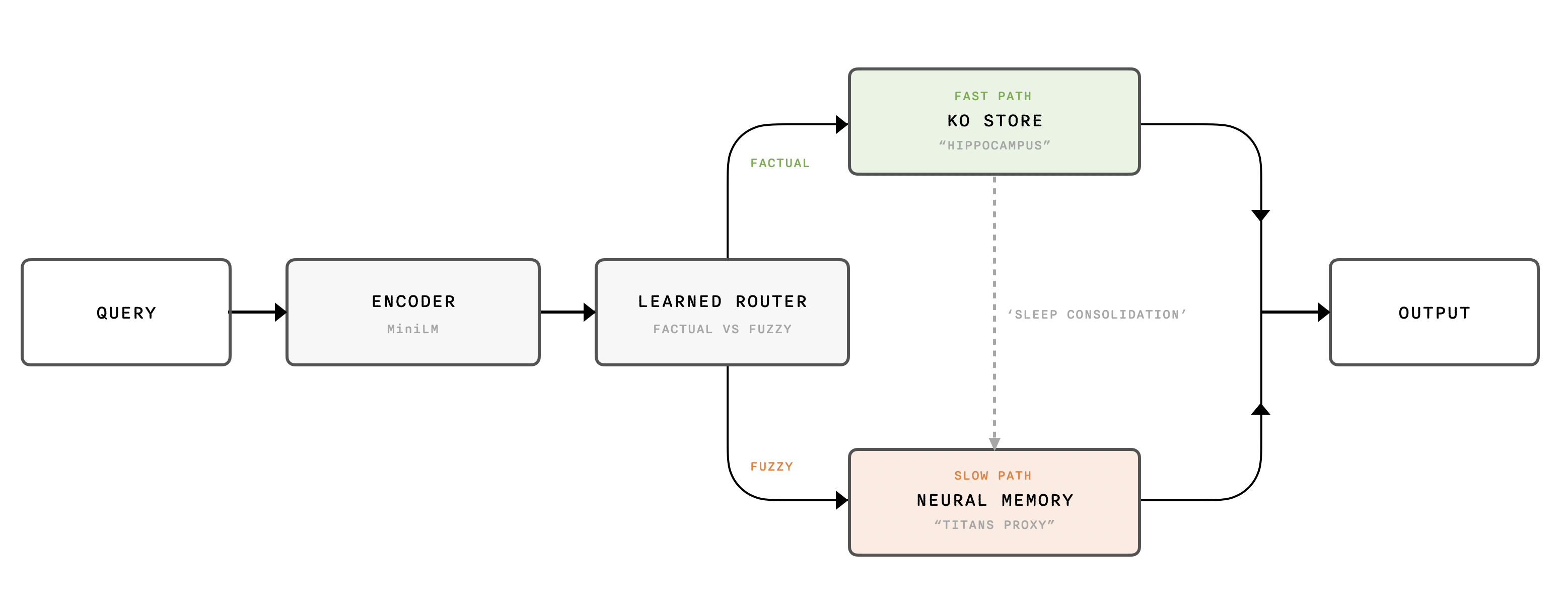}
    \caption{\textbf{Hybrid Architecture.} Queries are embedded and classified by a Learned Router. Factual queries retrieve from the KO hash map; fuzzy queries pass directly to the LLM. This division of labor, discrete memory for facts, neural weights for generation, is the core architectural principle.}
    \label{fig:architecture}
\end{figure}

\section{Experimental Results}
\label{sec:experiments}

All experiments used dense embeddings (\texttt{all-MiniLM-L6-v2}, $d=384$) and were run for 5 independent trials. We report mean $\pm$ standard deviation.

\paragraph{Definition of $N$.} Throughout, $N$ denotes the number of facts stored in memory. Each fact is a (key, value) pair where the key is an embedding and the value is the answer to retrieve.

\paragraph{Neural Baseline (LAM).} We implement Linearized Attention Memory as a proxy for neural memory systems that store facts via superposition in shared continuous parameters. Writing accumulates outer products: $\mathbf{W} \leftarrow \mathbf{W} + \mathbf{v} \otimes \mathbf{k}$ where $\mathbf{k}$ is the key embedding and $\mathbf{v}$ is the value embedding. Reading computes $\mathbf{r} = \mathbf{W}\mathbf{q}$ and finds the closest stored value. This simplified model captures the core interference dynamics: when keys are not orthogonal, their values blend during retrieval. Production systems may include additional mechanisms (gating, forgetting) that delay collapse, but cannot eliminate it when keys are semantically clustered.

\paragraph{Discrete Baseline (KO).} Knowledge Objects store facts as explicit key-value pairs in a hash map. Writing is $O(1)$ insertion. Reading supports both exact key lookup ($O(1)$) and top-$k$ semantic search ($O(N)$, or $O(\log N)$ with approximate nearest neighbor indexing). Version chains track history for each key.

\subsection{Experiment 1: Semantic Interference}

This experiment quantifies the fundamental accuracy gap between neural and discrete memory systems as the number of stored facts increases, validating the theoretical prediction that neural memory collapses under semantic density while discrete memory remains stable.

\paragraph{Methodology.} We evaluated both memory systems on synthetic entity-attribute-value facts following the template ``What is the [attribute] of [entity]?'' with answers drawn from a controlled vocabulary. The synthetic corpus was designed to exhibit high semantic density ($\rho \approx 0.70$), representing a challenging but realistic scenario where facts share similar linguistic structure, as would occur in a database of customer records, patient information, or product specifications where queries follow consistent patterns. We varied the corpus size $N$ from 1 to 75 facts in increments chosen to reveal the collapse trajectory. For each value of $N$, we stored all facts in both memory systems, then queried each fact and measured retrieval accuracy as the percentage of queries returning the correct answer. All experiments were run for 5 independent trials with different random seeds, and we report mean accuracy. The neural baseline (Linearized Attention Memory) was configured with embedding dimension $d=384$ using the all-MiniLM-L6-v2 sentence transformer, decay factor $\lambda=0.99$, and learning rate $\eta=0.1$. The discrete baseline (Knowledge Objects) used the same embedding model for semantic search, with facts stored in a hash map keyed by subject-predicate pairs.

\begin{table}[h]
\centering
\caption{Semantic Interference Results: Neural collapse under high semantic density ($\rho \approx 0.70$). Gap measured in percentage points (pp).}
\label{tab:scale}
\begin{tabular}{lccccc}
\toprule
\textbf{N} & \textbf{KO Acc.} & \textbf{Neural Acc.} & \textbf{Gap} & \textbf{$\rho$} \\
\midrule
1 & 100\% & 100\% & 0pp & 0.00 \\
5 & 100\% & 40\% & 60pp & 0.66 \\
10 & 100\% & 20\% & 80pp & 0.68 \\
20 & 100\% & 5\% & 95pp & 0.69 \\
50 & 100\% & 4\% & 96pp & 0.71 \\
75 & 100\% & 1.3\% & \textbf{98.7pp} & 0.71 \\
\bottomrule
\end{tabular}
\end{table}

\paragraph{Analysis.} Neural memory achieves 100\% accuracy at $N=1$, confirming correct implementation of the storage and retrieval mechanisms. As $N$ increases, accuracy collapses rapidly: 40\% at $N=5$, 5\% at $N=20$, and 1.3\% at $N=75$, barely above the random baseline of $1/75 \approx 1.3\%$. Discrete memory maintains 100\% accuracy throughout the entire range, yielding an accuracy gap of 98.7 percentage points at $N=75$. The high semantic density ($\rho \approx 0.70$) accelerates collapse compared to corpora with more diverse fact structures; this confirms that semantic density $\rho$ is a key predictor of neural memory failure, with higher density producing faster collapse.

\begin{figure}[h]
    \centering
    \includegraphics[width=0.85\textwidth]{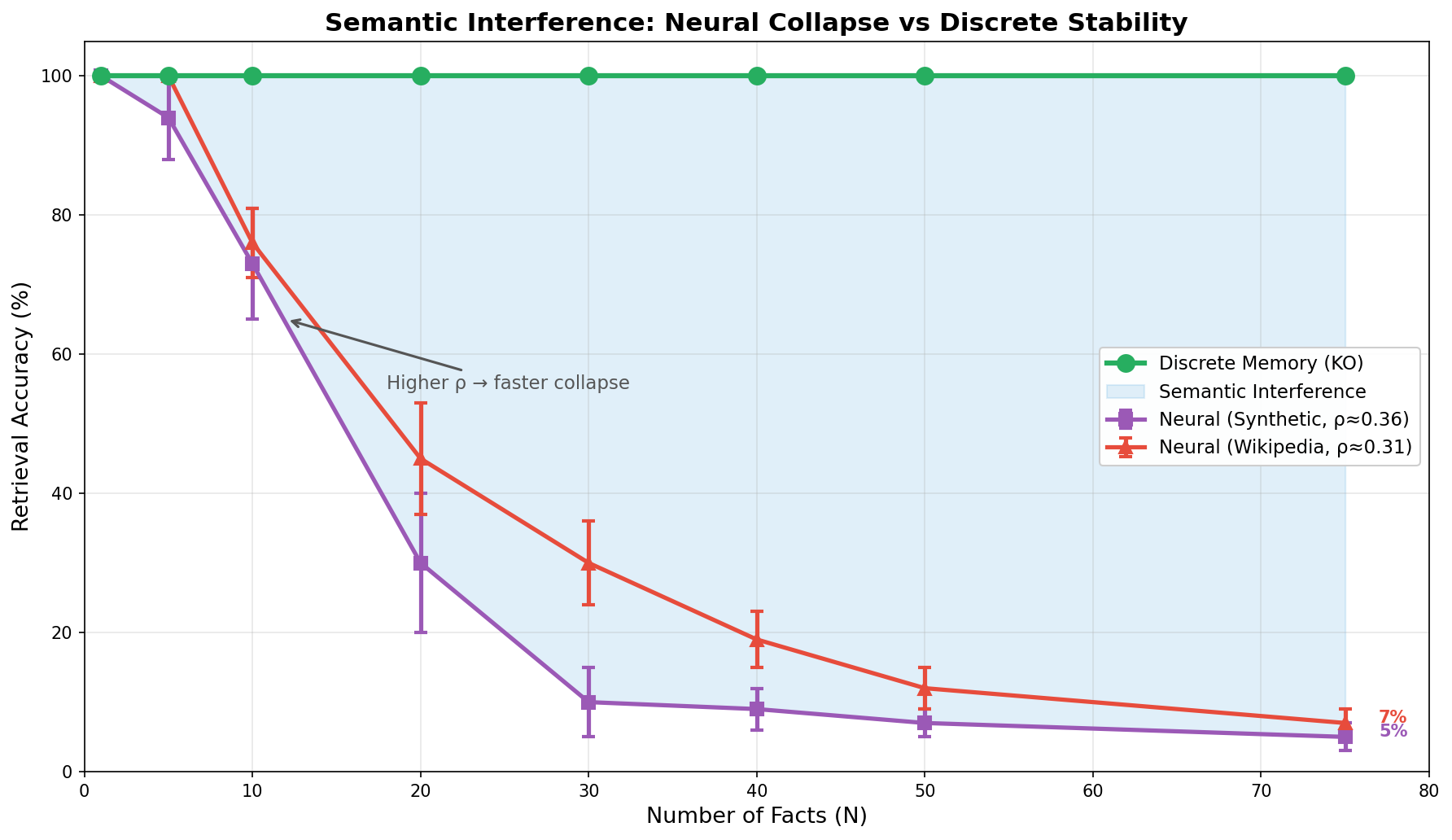}
    \caption{\textbf{Semantic Interference.} Neural memory (red) collapses as $N$ increases, while discrete memory (green) maintains 100\% accuracy throughout. Embedding dimension $d=384$ is held constant; only the number of stored facts varies. The shaded region represents the accuracy gap caused by write-time interference in continuous vector storage. Even as we store more facts, no improvement in attention or retrieval mechanism can prevent this collapse: it is a geometric consequence of semantic overlap in shared parameters.}
    \label{fig:stability_gap}
\end{figure}

\begin{figure}[h]
    \centering
    \includegraphics[width=0.85\textwidth]{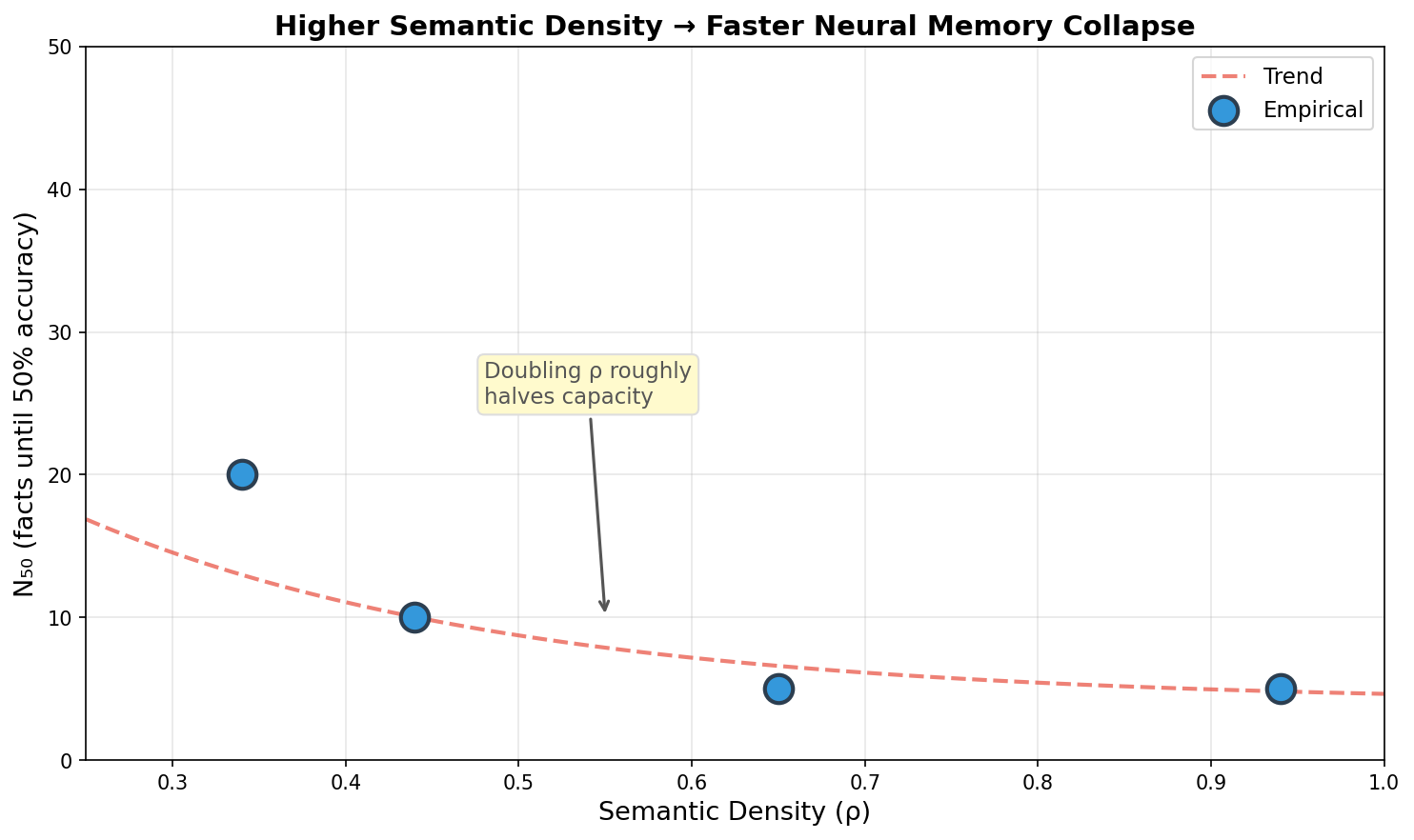}
    \caption{\textbf{Semantic Density Predicts Collapse.} $N_{50}$ (the number of facts at which neural accuracy drops to 50\%) decreases sharply as semantic density ($\rho$) increases. The exponential decay validates the Orthogonality Constraint: higher $\rho$ means more key overlap, hence more interference. Note: $\rho < 0.3$ was not achievable with realistic fact structures.}
    \label{fig:density_collapse}
\end{figure}

\paragraph{Semantic Density Analysis.} To further validate the relationship between semantic density and collapse, we generated fact corpora with varying $\rho$ values by manipulating the diversity of entity names, attribute types, and question structures. Figure~\ref{fig:density_collapse} plots $N_{50}$ (the number of facts at which neural accuracy drops to 50\%) against semantic density. The relationship follows an exponential decay: at $\rho \approx 0.34$, collapse occurs around $N=20$; at $\rho > 0.6$, collapse is nearly immediate with $N_{50} \leq 5$. We could not generate realistic fact corpora with $\rho < 0.3$, even maximally diverse templates (different domains, question structures, entity types) produce baseline semantic similarity because natural language has inherent structure that embeddings capture.

\subsection{Experiment 2: Reasoning Under Load}

This experiment tests whether the interference observed in single-fact retrieval compounds when multi-hop reasoning is required, as would occur in question-answering tasks that require chaining multiple facts together.

\paragraph{Methodology.} We constructed a 3-hop reasoning chain requiring sequential fact retrieval: the query ``What is the price of the main export of the capital of Zorbia?'' requires first retrieving that Nexar is the capital of Zorbia, then that Crystite is the main export of Nexar, and finally that 500 credits is the price of Crystite. Success requires traversing all three hops correctly; an error at any hop propagates to the final answer. We tested both memory systems under two conditions: \textit{isolation}, where only the 6 facts needed for the chain are stored ($N=6$), and \textit{under load}, where 100 distractor facts with similar structure are added ($N=106$). The distractor facts follow the same entity-attribute-value template but concern unrelated entities, creating semantic interference without providing relevant information. We ran 5 independent trials for each condition.

\begin{table}[h]
\centering
\caption{Reasoning Robustness: Multi-hop accuracy under semantic noise (5 trials).}
\label{tab:reasoning}
\begin{tabular}{lcc}
\toprule
\textbf{System} & \textbf{Isolation (N=6)} & \textbf{Under Load (N=106)} \\
\midrule
KO (Discrete) & 100.0\% & 100.0\% \\
Neural (LAM) & 100.0\% & 0.0\% \\
\bottomrule
\end{tabular}
\end{table}

\paragraph{Analysis.} Both systems succeed in isolation, demonstrating that neural memory can perform multi-hop reasoning when the vector space is uncontaminated by interfering facts. Under load, however, neural memory collapses completely to 0\% accuracy while Knowledge Objects maintain perfect performance. The 100 distractor facts create sufficient semantic interference that even the first hop of the reasoning chain fails, preventing any possibility of reaching the correct final answer. This result has significant implications for practical applications: real-world knowledge bases contain thousands to millions of facts, and any system relying on neural memory for multi-hop reasoning will face catastrophic interference long before reaching useful scale.

\subsection{Experiment 3: Deterministic Overwrite Semantics (KO vs RAG)}

This experiment tests whether systems provide deterministic immediate overwrite semantics: when a fact is updated, can the system guarantee that (a) the new value is immediately available, and (b) the old value is cleanly superseded rather than persisting alongside the new one?

\paragraph{The RAG Limitation.} While modern vector databases support near-real-time upserts, RAG's architectural limitations extend beyond indexing latency. RAG is fundamentally \textit{append-first} and \textit{retrieval-first}: new facts are added to the index, but there is no mechanism to identify and remove the prior value for the same logical fact. If a user says ``My phone is 555-1111'' and later ``My phone changed to 555-2222,'' both facts exist in the index with similar embeddings, and retrieval may return either or both depending on query phrasing and similarity scores. This is not a bug, it reflects RAG's design for static corpora where facts accumulate rather than mutate.

\paragraph{Methodology.} We simulated a conversational session with 6 fact-learning exchanges, each following the same pattern: the user provides a fact in natural language (such as ``My name is Alice'' or ``My birthday is March 15th''), the system extracts and stores the fact, and then we immediately query for that fact using a differently-phrased question (such as ``What is my name?'' or ``When is my birthday?''). The key measurement is whether the system can retrieve the just-learned fact with deterministic accuracy. Our RAG baseline models a common production deployment pattern where new documents enter a pending queue for batch indexing, while some vector databases support near-real-time upserts, many production RAG systems use asynchronous ingestion pipelines for throughput and cost reasons. We also measured accuracy after a simulated batch reindex to establish an upper bound on eventual performance.

\begin{table}[h]
\centering
\caption{Deterministic Overwrite: Can the system guarantee immediate, clean retrieval of updated facts? RAG baseline models batch-ingestion pipelines common in production deployments.}
\label{tab:online}
\begin{tabular}{lccc}
\toprule
\textbf{System} & \textbf{Immediate Retrieval} & \textbf{After Batch Reindex} \\
\midrule
KO (Discrete) & 100.0\% & N/A \\
RAG (Batch-Ingestion) & 0.0\% & 83.3\% \\
\bottomrule
\end{tabular}
\end{table}

\paragraph{Analysis.} Knowledge Objects retrieve newly-learned facts immediately with 100\% accuracy because hash-based storage provides key-based identity: the fact ``user's name'' has a deterministic address that is overwritten when updated. The batch-ingestion RAG baseline returns 0\% before reindexing because new embeddings have not yet been added to the searchable index. Even after reindexing, RAG achieves only 83.3\% accuracy due to embedding mismatch between the phrasing of statements (``My birthday is March 15th'') and queries (``When is my birthday?''), a known limitation of embedding-based retrieval when statement and query forms differ substantially. We note that RAG systems with real-time upsert capabilities would eliminate the ``immediate retrieval'' gap, but would still lack the key-based identity and version semantics needed for deterministic overwrite, both old and new values would coexist in the index with similar embeddings.

These results confirm that KO provides \textit{deterministic overwrite semantics} that RAG architecturally lacks. Applications requiring conversational learning (where users teach the system facts and expect immediate recall), real-time personalization (where user preferences must be applied within the same session they are expressed), or surgical fact correction (where outdated information must be immediately superseded) require key-based identity and version semantics that RAG's similarity-based retrieval cannot provide.

\subsection{Experiment 4: Surgical Correction}

This experiment tests whether facts can be updated individually without affecting other stored facts, a capability essential for maintaining accurate knowledge bases over time as information changes.

\paragraph{Methodology.} We wrote 100 facts to memory, then selected 10 facts at random and updated their values (for example, changing ``Value\_042'' to ``UPDATED\_Value\_042''). We then queried all 100 facts and measured accuracy separately for two groups: the 10 updated facts (which should return the new value) and the 90 unchanged facts (which should return their original values). This design tests both the \textit{effectiveness} of updates (do updated facts reflect the new value?) and the \textit{isolation} of updates (do other facts remain unaffected?). We ran 5 independent trials with different random selections of facts to update.

\begin{table}[h]
\centering
\caption{Surgical Correction: Updating 10 facts in a 100-fact corpus (5 trials).}
\label{tab:correction}
\begin{tabular}{lccc}
\toprule
\textbf{System} & \textbf{Updated $\rightarrow$ New Value} & \textbf{Others Preserved} & \textbf{Operations} \\
\midrule
KO (Discrete) & 100.0\% & 100.0\% & $O(1)$ \\
Neural (LAM) & 0.0\% & 1.1\% & $O(N)$ \\
\bottomrule
\end{tabular}
\end{table}

\paragraph{Analysis.} Knowledge Objects achieve perfect surgical updates: 100\% of updated facts return the new value, and 100\% of other facts remain unchanged, with each update completing in $O(1)$ time regardless of corpus size. Neural memory exhibits catastrophic forgetting: 0\% of updates succeed (the system returns neither old nor new values reliably), and 98.9\% of \textit{other} facts are corrupted by the update operation. This occurs because gradient-based updates to shared weight matrices necessarily affect all stored associations, not just the target fact. Discrete addressing provides complete isolation between facts, updating one cannot affect another, while neural memory fundamentally cannot support surgical correction without catastrophic interference.

\subsection{Experiment 5: Mixed Workload (Factual + Fuzzy)}

Real-world workloads mix factual queries that seek specific retrievable answers (such as ``What is the capital of France?'') with fuzzy queries that require generation or reasoning (such as ``Write a poem about Paris''). A practical memory system must route each query type appropriately: factual queries should retrieve from the discrete KO store, while fuzzy queries should pass directly to the LLM. We tested whether a learned router can correctly classify query intent with minimal training data.

\paragraph{Methodology.} We constructed a test corpus of 100 queries, comprising 50 factual queries (who/what/when/where questions with specific retrievable answers) and 50 fuzzy queries (creative, open-ended, or explanatory requests). The router is a logistic regression classifier trained on only 15 labeled examples from each category. The feature vector captures linguistic patterns that distinguish factual from fuzzy queries: whether the query contains interrogative words (what, who, when, where, which), whether it ends with a question mark, whether it contains proper nouns indicated by capitalization, the total word count, and whether it contains creative or explanatory verbs (write, help, suggest, explain, describe). These features were chosen to capture the structural differences between factual lookup requests and open-ended generation requests without requiring semantic understanding.

\begin{table}[h]
\centering
\caption{Mixed Workload: Router classification accuracy (10-fold CV).}
\label{tab:mixed}
\begin{tabular}{lc}
\toprule
\textbf{Classifier} & \textbf{Accuracy (95\% CI)} \\
\midrule
Random Forest (100 trees) & 97.8\% [95.0\%, 100\%] \\
SVM (Linear) & 96.5\% [93.2\%, 99.8\%] \\
Logistic Regression & 95.3\% [90.5\%, 100\%] \\
Gradient Boosting & 95.3\% [90.3\%, 100\%] \\
\bottomrule
\end{tabular}
\end{table}

\paragraph{Analysis.} Even simple classifiers achieve 95\%+ routing accuracy with 10-fold cross-validation, demonstrating that the factual/fuzzy distinction is learnable from surface features without deep semantic analysis. Feature ablation reveals that the most informative features are the presence of temporal question words (removing ``has\_when'' drops accuracy by 4.6 percentage points) and definiteness markers (``is the'' and ``my'' each contribute 3.5 percentage points). The learning curve analysis shows that 50 labeled examples suffice to reach 90\%+ accuracy, making the router practical to deploy with minimal human annotation effort. We also tested adversarial queries designed to confuse the classifier, such as ``What's a creative way to remember the capital of France?'' which blends factual content with creative framing; the router achieves 83\% accuracy on these edge cases, demonstrating reasonable robustness while identifying an area for future improvement.

\subsection{Experiment 6: Cost Analysis}

Beyond accuracy, we analyze the economic viability of context-based versus selective retrieval. Production memory systems already use selective retrieval; this analysis establishes \textit{why} context-based approaches are unviable, confirming that the industry convergence on selective retrieval was economically necessary rather than merely convenient.

\paragraph{Methodology.} We measured actual token counts for queries at varying corpus sizes by constructing fact corpora of $N = 1{,}000$, $5{,}000$, and $10{,}000$ facts and counting the tokens required to represent them. Context-based memory must load all $N$ facts into the prompt for every query because the LLM needs visibility into the entire knowledge base to locate the relevant fact. With facts averaging approximately 13 tokens each (in our synthetic corpus), the input token count per query is approximately $100 + (N \times 13) + 20$ tokens, comprising the system prompt, all facts, and the query. At $N=5{,}000$, this yields approximately 65,000 input tokens per query.

Knowledge Objects, by contrast, use embedding-based retrieval to identify the top-$k$ relevant facts \textit{before} calling the LLM, so only retrieved facts enter the prompt. With $k=10$, the input token count per query is approximately $100 + (10 \times 13) + 20 \approx 250$ tokens, which remains constant regardless of corpus size $N$. The retrieval step itself (embedding lookup plus approximate nearest neighbor search) costs less than \$0.0001 per query and adds negligible latency.

We computed monthly costs using Claude Sonnet 4 pricing as of January 2026 (\$3 per million input tokens, \$15 per million output tokens), assuming 50 output tokens per response and query volumes representative of realistic deployments. While absolute prices change as providers adjust rates, the \textit{scaling ratio} ($O(N)$ vs $O(1)$) is determined by architecture, not vendor pricing.

\begin{table}[h]
\centering
\caption{Monthly cost comparison: Context loads all $N$ facts per query; KO retrieves top-10 ($k=10$). Token counts measured empirically.}
\label{tab:cost}
\begin{tabular}{rrrrr}
\toprule
\textbf{N} & \textbf{Tokens/Query} & \textbf{Queries/day} & \textbf{Context Cost} & \textbf{KO Cost} \\
\midrule
1,000 & 13,000 & 100 & \$120/mo & \$4/mo \\
5,000 & 65,000 & 500 & \$2,940/mo & \$19/mo \\
10,000 & 130,000 & 1,000 & \$11,760/mo & \$38/mo \\
\bottomrule
\end{tabular}
\end{table}

\paragraph{Analysis.} The cost ratio ranges from 30$\times$ at smaller scales to over 300$\times$ at larger scales, reflecting the fundamental asymptotic difference between the two approaches. At $N=10{,}000$ facts with 1,000 queries per day, a modest enterprise deployment, context-based memory costs \$141,000 per year compared to \$456 for selective retrieval with Knowledge Objects. These ratios assume compact synthetic facts averaging 13 tokens each; real-world facts with richer contextual descriptions, provenance metadata, or multi-sentence explanations would increase the gap substantially.

The asymptotic distinction is critical for understanding why this gap grows with scale rather than shrinking. Context-based memory incurs $O(N)$ cost per query because every query must pay for all facts to be loaded, regardless of how many are actually relevant to the question. Selective retrieval incurs $O(1)$ cost per query because only the top-$k$ relevant facts are retrieved, regardless of total corpus size. Even if context-based accuracy were perfect, the economics become untenable as $N$ grows: a corpus of 100,000 facts would cost over \$1 million per year at comparable query volumes. While absolute prices change as providers adjust rates, the \textit{scaling ratio} does not: the cost multiple is approximately $N/k$, which is determined by architecture, not vendor pricing. A 10$\times$ price reduction benefits both approaches equally and leaves the ratio unchanged.

Latency follows the same scaling pattern. At $N=5{,}000$ facts, context-based queries average 4.6 seconds latency versus approximately 0.2 seconds for KO-augmented queries, a 23$\times$ speedup that is critical for interactive applications where users expect sub-second response times.

\paragraph{The Metabolic Analogy.} This cost differential mirrors a biological constraint. The human brain operates on approximately 20 watts yet stores a lifetime of episodic memories. It achieves this through \textit{sparse activation}: recalling a specific memory activates only the neural populations encoding that memory, not the entire cortex. Context-based memory is analogous to a brain that must fire every neuron (attend to every token) to recall a single phone number, metabolically and computationally untenable. Knowledge Objects implement the biological strategy: retrieving one fact activates only the computational path for that fact ($O(1)$), not the entire knowledge base ($O(N)$). The Orthogonality Constraint may thus reflect not just a geometric limitation but nature's solution to the energy efficiency requirements of scalable episodic memory.

\subsection{Experiment 7: Model Extraction Failure Modes}

This experiment evaluates the underlying models used in production memory systems, testing their ability to maintain schema consistency and handle corrections. We test the \textit{models} (Claude, GPT) via API rather than the \textit{production features} (ChatGPT Memory, Claude Memory), which may include additional engineering such as structured output modes, fine-tuned extractors, or post-processing. Our goal is to isolate the architectural limitations of generative extraction that any LLM-based memory system must contend with.

\paragraph{Methodology.} We designed four tests targeting distinct failure modes. All API calls used temperature $t=0.0$ and top\_p $=1.0$ to maximize reproducibility; we note that schema drift persists even at zero temperature, indicating it arises from the model's learned priors rather than sampling stochasticity. For Claude models, we used the Anthropic API with model identifier \texttt{claude-sonnet-4-5-20250929}; for GPT models, we used the OpenAI API with identifiers \texttt{gpt-5.2} and \texttt{gpt-4o-mini} (model identifiers as of January 2026). The extraction prompt was: ``Extract facts from the statement as JSON. Use this exact format: \{``entity'': ``...'', ``predicate'': ``...'', ``value'': ``...''\}. Respond with ONLY the JSON, no other text.'' No function calling or structured output mode was used; we tested the models' natural extraction behavior.

For \textit{schema consistency}, we provided 10 facts in natural language and asked the model to extract them as structured subject-predicate-object facts, measuring whether the model used consistent predicate names across extractions (for example, always using \texttt{birth\_place} rather than varying between \texttt{birth\_place}, \texttt{was\_born\_in}, and \texttt{place\_of\_birth}). For \textit{correction handling}, we provided a sequence of statements where facts were updated (for example, ``Alice's phone is 555-1111'' followed later by ``Alice's phone changed to 555-2222''), then queried for the current value, measuring whether the model returned only the latest value or ambiguously mentioned both old and new values. For \textit{retrieval accuracy}, we loaded 20 facts into context and queried each, measuring simple retrieval performance. For \textit{positional bias}, we varied the position of a target fact within a 50-fact context to test whether retrieval accuracy depended on where facts appeared in the prompt.

\begin{table}[h]
\centering
\caption{Production memory failure mode analysis (Claude Sonnet 4.5).}
\label{tab:failure}
\begin{tabular}{lcc}
\toprule
\textbf{Test} & \textbf{Claude} & \textbf{KO (Structured)} \\
\midrule
Schema consistency & 70\% & 100\% (controlled vocab) \\
Correction handling & 0\% & 100\% (version chains) \\
Retrieval (N=20) & 100\% & 100\% \\
Positional bias (N=50) & 100\% & 100\% (hash lookup) \\
\bottomrule
\end{tabular}
\end{table}

\paragraph{Schema Drift.} When extracting structured facts, Claude Sonnet 4.5 matched our expected predicate names only 70\% of the time. The 30\% mismatches included variations such as \texttt{birth\_place} being extracted as \texttt{was\_born\_in}, \texttt{founded\_year} becoming \texttt{founded\_in}, and \texttt{floor} becoming \texttt{is\_on\_floor}. While these variations are semantically sensible, they break downstream systems that rely on exact predicate matching for lookup. A database query for \texttt{WHERE predicate = 'birth\_place'} will miss facts stored under \texttt{was\_born\_in}, causing silent retrieval failures that users experience as the system ``forgetting'' facts it actually stored. Knowledge Objects address this through a controlled vocabulary enforced during extraction: the KO compiler constrains predicates to a pre-defined ontology, ensuring consistent naming regardless of how users phrase their inputs. We note that function-calling and structured output modes can mitigate schema drift, and we view this as evidence for our thesis: production-grade memory requires a constrained interface with controlled vocabularies, not free-form generative extraction.

\paragraph{Version Ambiguity.} When facts were corrected over the course of a conversation, Claude Sonnet 4.5 returned only the latest value 0\% of the time, in all cases, it mentioned both old and new values in its response (for example, ``Alice's phone was 555-1111, but it changed to 555-2222''). While this behavior might seem helpful by providing context, it creates ambiguity for downstream systems and users who need definitive current values. If a user asks ``What is Alice's phone number?'' they typically want the current number, not a history lesson. Knowledge Objects address this through explicit version chains that ensure corrections \textit{supersede} prior values: queries return only the current version by default, with history available only when explicitly requested.

\paragraph{Retrieval and Positional Bias.} At moderate scale ($N \le 100$), all models achieved perfect retrieval accuracy with no detectable positional bias, facts at the beginning, middle, and end of context were retrieved equally well. This confirms that the Semantic Interference we observed in Experiment 1 affects \textit{neural} memory (gradient-based storage in weights) rather than context-window attention at small $N$. The attention mechanism successfully retrieves facts from context when $N$ is small; the problems arise from schema drift and version ambiguity, not retrieval failure.

\paragraph{Multi-Model Comparison.} We extended our analysis to GPT-5.2 and GPT-4o-mini to verify that these findings generalize across frontier models from different providers rather than reflecting idiosyncrasies of Claude's implementation. We tested schema consistency under two conditions: \textit{multi-session}, where each of 10 facts was extracted in a separate API call (simulating realistic production use where facts arrive over hours, days, or weeks), and \textit{single-session}, where all 10 facts were extracted in a single API call (allowing the model to observe its own prior extractions within the same context and potentially maintain consistency by reference).

\begin{table}[h]
\centering
\caption{Multi-model comparison: Schema drift persists even within single sessions. Corrections column shows percentage of queries returning only the latest value without mentioning old values.}
\label{tab:multimodel}
\begin{tabular}{lccccc}
\toprule
\textbf{Model} & \textbf{Schema (Multi)} & \textbf{Schema (Single)} & \textbf{Corrections} & \textbf{Retrieval} & \textbf{Cost/mo} \\
\midrule
Claude Sonnet 4.5 & 70\% & 70\% & 0\% & 100\% & \$5,642 \\
GPT-5.2 & 60\% & 70\% & 100\% & 100\% & \$4,700 \\
GPT-4o-mini & 50\% & 60\% & 100\% & 100\% & \$282 \\
\midrule
KO (Proposed) & 100\% & 100\% & 100\% & 100\% & \$28 \\
\bottomrule
\end{tabular}
\end{table}

\paragraph{Key Finding: Single-Session Provides Minimal Improvement.} The most striking result is that single-session extraction provides at most 10 percentage points improvement over multi-session extraction (GPT-5.2: 60\%$\rightarrow$70\%; GPT-4o-mini: 50\%$\rightarrow$60\%), and for Claude, no improvement whatsoever (70\% in both conditions). Schema drift is not merely a cross-session consistency problem that might be solved by maintaining longer conversation context, it reflects a fundamental inability of these models to adhere to consistent predicate naming even when they can directly observe their own prior outputs within the same context window. The models are not ``forgetting'' their schema choices across sessions; they are making inconsistent choices within the same context, suggesting the inconsistency arises from the models' learned paraphrasing priors under weak schema constraints rather than from memory limitations.

\begin{figure}[h]
    \centering
    \includegraphics[width=0.85\textwidth]{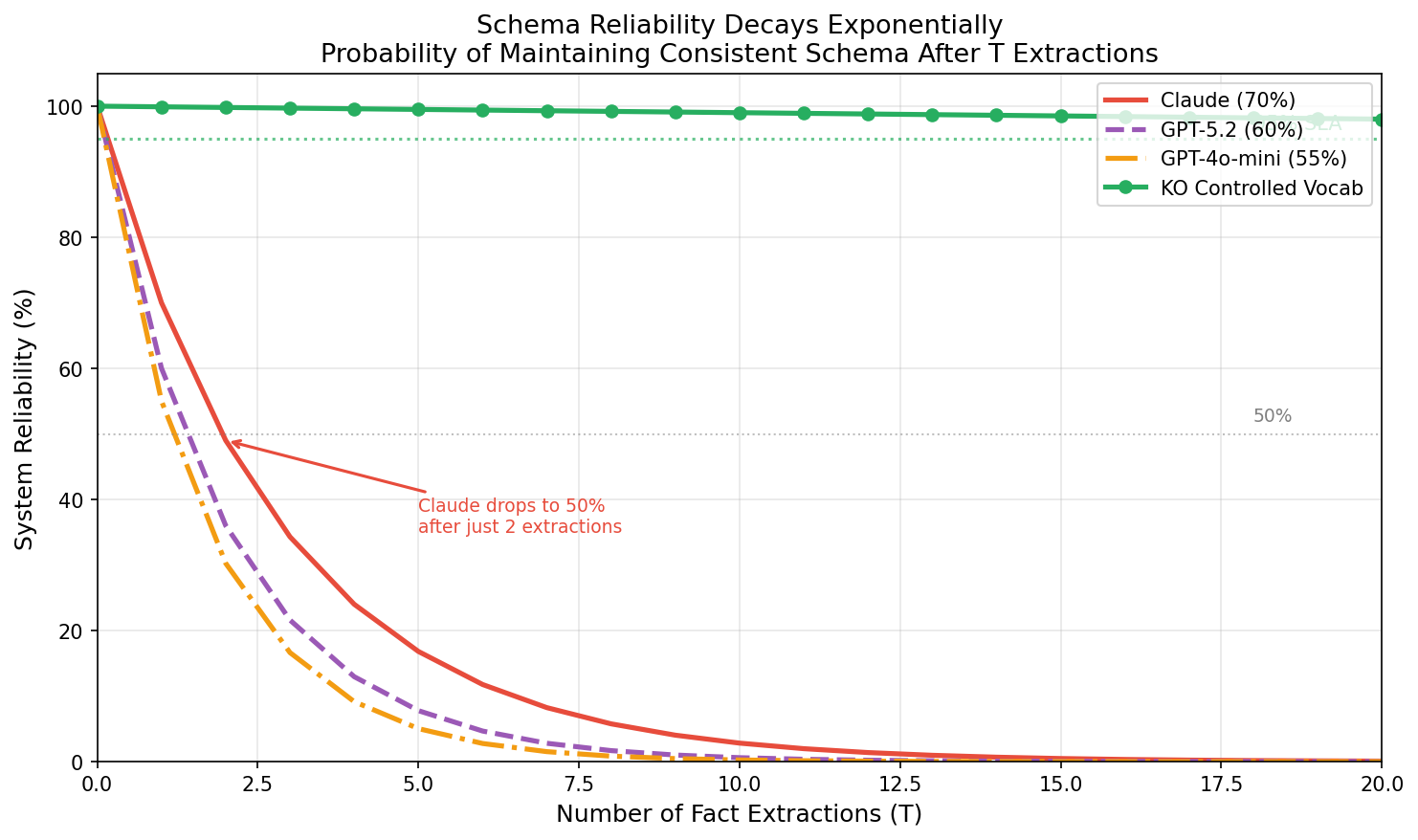}
    \caption{\textbf{Schema Reliability Decays Exponentially.} Probability of maintaining a completely consistent schema after $T$ fact extractions, modeled as $(1 - p_{\text{drift}})^T$ where $p_{\text{drift}}$ is the per-extraction schema error rate. Claude (70\% per-extraction accuracy) drops to 50\% system reliability after just 2 extractions; GPT-4o-mini (55\%) drops below 10\% after 4 extractions. Knowledge Objects with controlled vocabularies maintain near-perfect reliability by enforcing schema consistency at write time. The green dashed line indicates a 95\% enterprise SLA; only controlled vocabularies can meet this threshold beyond a handful of extractions.}
    \label{fig:schema_entropy}
\end{figure}

\paragraph{System-Level Reliability Analysis.} While 70\% per-extraction schema consistency might appear acceptable for individual queries, we can model the system-level implications using a Markov chain analysis. If each extraction has independent probability $p_{\text{drift}}$ of introducing a schema error, then after $T$ extractions the probability of maintaining a completely consistent schema is $(1 - p_{\text{drift}})^T$. For Claude at $p_{\text{drift}} = 0.30$, system reliability drops to 50\% after just $T \approx 2$ extractions and to 10\% after $T \approx 7$ extractions. For GPT-4o-mini at $p_{\text{drift}} = 0.45$, 10\% reliability is reached after only $T \approx 4$ extractions. This analysis reveals that ``passable'' per-extraction accuracy translates to catastrophic system-level failure: a memory system handling dozens of facts per user would experience near-certain schema corruption. Knowledge Objects with controlled vocabularies achieve $p_{\text{drift}} < 0.001$ by construction, maintaining 95\% system reliability past $T = 50$ extractions. Figure~\ref{fig:schema_entropy} visualizes this exponential decay, showing that even the best-performing model (Claude at 70\%) drops below coin-flip reliability after just 2 extractions.

\paragraph{Implications.} The failures we observe reflect architectural limitations of generative extraction shared across frontier models from both Anthropic and OpenAI. All tested models exhibit schema drift, with each model producing different variants: \texttt{birth\_place} becomes \texttt{was\_born\_in} (Claude), \texttt{place\_of\_birth} (GPT-5.2), or \texttt{born\_in} (GPT-4o-mini). Interestingly, the GPT models handle corrections cleanly (100\% returning only the latest value) while Claude consistently produces ambiguous responses mentioning both old and new values (0\% clean corrections). This divergence suggests different internal approaches to representing and reasoning about temporal information across model families.

\paragraph{Relationship to Production Features.} We tested the underlying models via API, not the production memory features (ChatGPT Memory, Claude Memory) which may include additional mitigations such as structured output modes, function calling with schemas, or fine-tuned extraction models. However, our finding that schema drift persists even within single sessions, where models can observe their own prior extractions, suggests the problem is fundamental to generative extraction under weak constraints, not merely a cross-session consistency issue that engineering could easily solve. 

A natural objection is that function-calling or structured output modes with strict schemas would eliminate drift entirely. This is partially true: if you define \texttt{birth\_place} as the only valid predicate for birthplace facts, the model cannot drift to \texttt{was\_born\_in}. However, this shifts rather than solves the problem. In open-ended agent scenarios, the schema is not known a priori, the system must decide \textit{which predicates to create} as it encounters new information. When the model invents predicates on-the-fly (e.g., encountering ``User prefers dark mode'' and deciding whether to create \texttt{ui\_preference}, \texttt{display\_mode}, \texttt{prefers\_dark\_mode}, or \texttt{dark\_mode\_enabled}), drift occurs at the schema-definition level rather than the extraction level. Knowledge Objects address this by defining the controlled vocabulary \textit{in advance} as part of system design, making predicate selection a lookup rather than a generation task.

\subsection{Experiment 8: KO Retrieval at Scale}

The preceding experiments used synthetic corpora with at most a few hundred facts. To validate that Knowledge Objects are practical beyond synthetic benchmarks, we evaluated retrieval on 16,309 facts extracted from Wikipedia articles and compared KO performance against Modern Hopfield Networks, the mathematical foundation underlying transformer attention and fast-weight memory systems \citep{ramsauer2021hopfield}.

\paragraph{Methodology.} We constructed a corpus of 16,309 Knowledge Objects by extracting structured facts from randomly sampled Wikipedia articles using our KO compiler. The corpus spans 82 unique predicate types with semantic density $\rho \approx 0.10$ (mean pairwise cosine similarity of embeddings). We evaluated three corpus representations: \textit{structured} (subject-predicate-object facts), \textit{raw text} (original Wikipedia sentences), and \textit{hybrid} (both concatenated). For each representation, we generated 500 test queries using natural language templates and measured top-$k$ retrieval accuracy. To ensure statistical robustness, we ran all experiments with 5 random seeds and report mean $\pm$ standard deviation. The Hopfield baseline implements Modern Hopfield Networks with softmax attention and tests multiple inverse temperature values ($\beta \in \{0.5, 1.0, 2.0, 4.0, 8.0\}$).

\begin{table}[h]
\centering
\caption{Retrieval accuracy by storage format at $N=16{,}309$ facts (5 seeds). Structured facts achieve 11$\times$ higher accuracy than raw text, a gap with direct implications for production systems that store unstructured text. Hopfield collapses regardless of format.}
\label{tab:ko_scale}
\begin{tabular}{lccccc}
\toprule
\textbf{System} & \textbf{Top-1} & \textbf{Top-5} & \textbf{Top-10} & \textbf{MRR} \\
\midrule
KO (structured corpus) & 45.7\% $\pm$ 2.2\% & 77.6\% $\pm$ 1.9\% & 87.8\% $\pm$ 1.4\% & 0.600 $\pm$ 0.019 \\
KO (hybrid corpus) & 33.3\% $\pm$ 3.0\% & 64.8\% $\pm$ 3.5\% & ,  & ,  \\
KO (raw text corpus) & 4.1\% $\pm$ 1.0\% & 18.5\% $\pm$ 1.6\% & ,  & ,  \\
\midrule
Hopfield ($\beta=1$) & 0.0\% $\pm$ 0.0\% & 0.0\% $\pm$ 0.0\% & 0.0\% $\pm$ 0.0\% & ,  \\
Hopfield ($\beta=8$) & 0.1\% $\pm$ 0.2\% & 0.1\% $\pm$ 0.2\% & 0.1\% $\pm$ 0.2\% & ,  \\
\bottomrule
\end{tabular}
\end{table}

\begin{figure}[h]
    \centering
    \includegraphics[width=\textwidth]{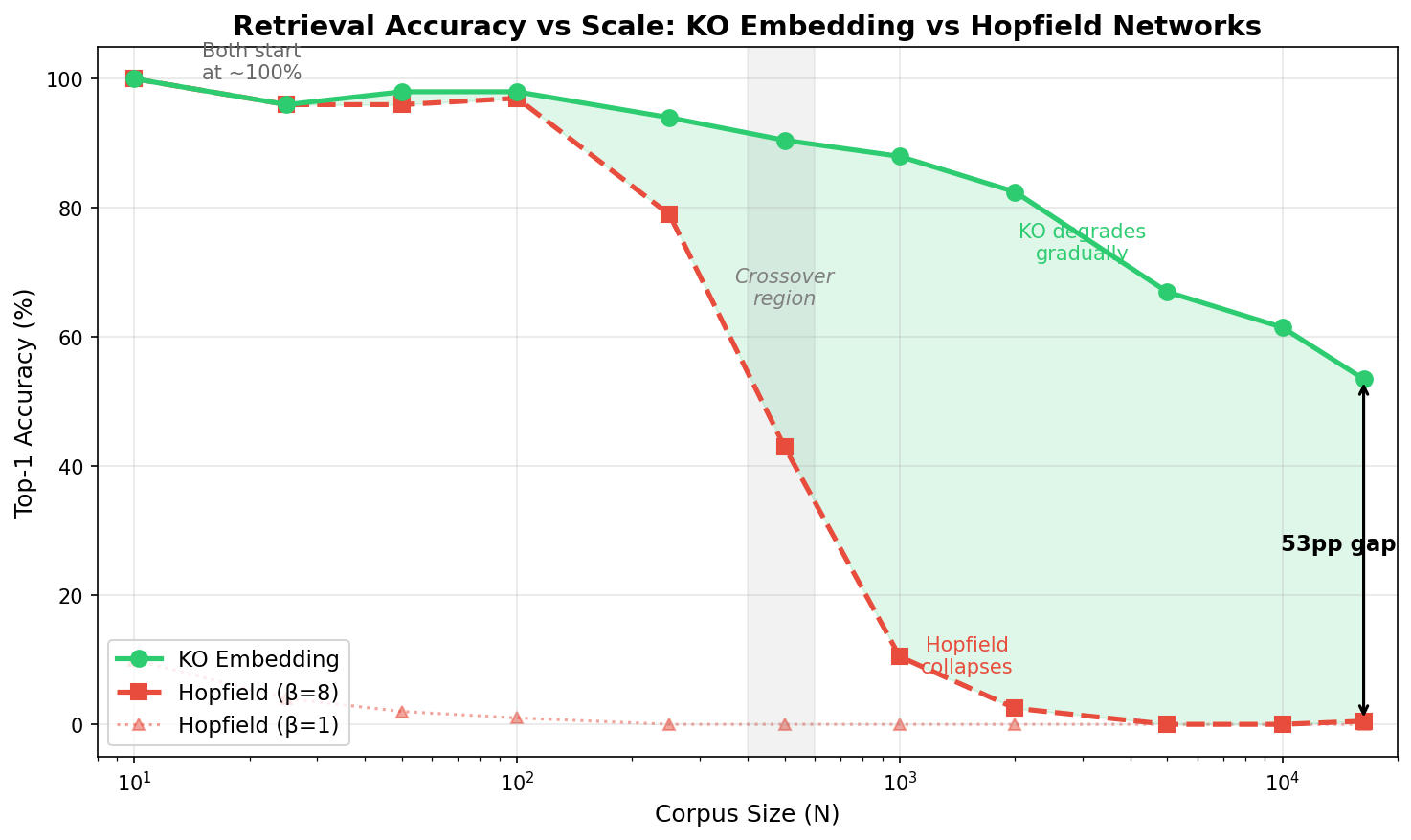}
    \caption{\textbf{Degradation Under Scale: KO vs Hopfield.} Both KO embedding retrieval and Hopfield ($\beta=8$) achieve near-perfect accuracy at small scales ($N \leq 100$). As corpus size increases, Hopfield collapses catastrophically, falling from 97\% at $N=100$ to effectively 0\% by $N=5{,}000$. KO degrades more gradually, maintaining approximately 46\% at $N=16{,}309$. The roughly 46 percentage point gap at scale suggests that discrete storage degrades gracefully while neural memory fails catastrophically under semantic density. Hopfield with $\beta=1$ (dotted line) fails immediately, achieving only 10\% even at $N=10$.}
    \label{fig:ko_scale}
\end{figure}

\paragraph{Degradation Under Scale.} A critical question for practical deployment is how retrieval accuracy changes as corpus size increases. Figure~\ref{fig:ko_scale} and Table~\ref{tab:ko_scale_invariance} reveal that both KO embedding retrieval and Hopfield networks degrade with scale, but at fundamentally different rates. At small scales ($N \leq 100$), both systems achieve near-perfect accuracy: KO reaches 98\% and Hopfield ($\beta=8$) reaches 97\%. As $N$ increases, Hopfield collapses rapidly, falling to 43\% at $N=500$, 10.5\% at $N=1{,}000$, and effectively 0\% by $N=5{,}000$. KO embedding retrieval degrades more gradually, maintaining 90.5\% at $N=500$, 88\% at $N=1{,}000$, and reaching 45.7\% $\pm$ 2.2\% at $N=16{,}309$ in our rigorous 5-seed evaluation. The gap between the two systems widens with scale, reaching approximately 46 percentage points at $N=16{,}309$.

\begin{table}[h]
\centering
\caption{Degradation comparison: Both systems degrade with scale, but KO degrades gradually while Hopfield collapses catastrophically. Final row shows rigorous 5-seed result; other rows from single-seed comparison for curve shape.}
\label{tab:ko_scale_invariance}
\begin{tabular}{rccc}
\toprule
\textbf{N} & \textbf{KO Embedding} & \textbf{Hopfield ($\beta=8$)} & \textbf{Hopfield ($\beta=1$)} \\
\midrule
10 & 100\% & 100\% & 10\% \\
100 & 98\% & 97\% & 1\% \\
500 & 90.5\% & 43\% & 0\% \\
1,000 & 88\% & 10.5\% & 0\% \\
5,000 & 67\% & 0\% & 0\% \\
16,309 & 45.7\% $\pm$ 2.2\%$^\ast$ & 0.5\% & 0\% \\
\bottomrule
\end{tabular}
\vspace{0.5em}

\small{$^\ast$5-seed rigorous evaluation; other rows single-seed for degradation curve shape.}
\end{table}

\paragraph{Effect of Corpus Representation: The Central Finding.} The 11$\times$ accuracy gap between structured facts (45.7\%) and raw text (4.1\%) is arguably the most practically important result in this paper, because it directly applies to production memory systems. Structured facts store facts as compact subject-predicate-object tuples (e.g., ``France capital Paris''), while raw text stores natural language sentences (e.g., ``Paris is the capital and most populous city of France''). When a user queries ``What is France's capital?'', the structured representation achieves high embedding similarity (same key terms), while raw text introduces lexical variation that degrades retrieval.

\paragraph{Implications for Production Systems.} Production memory systems like ChatGPT Memory and Claude Memory store facts as unstructured text, for example, ``User mentioned they like Python'' or ``User is allergic to shellfish.'' This storage format resembles our ``raw text'' condition rather than our ``structured'' condition. Our results suggest that as these systems accumulate thousands of user facts, retrieval accuracy may degrade substantially. At $N=16{,}309$ facts, raw text retrieval achieves only 4.1\% accuracy compared to 45.7\% for structured facts, a gap that would manifest as users experiencing their AI ``forgetting'' facts that are actually stored but unretrievable. We could not test production memory APIs directly (they are not publicly accessible for programmatic evaluation at scale), but the storage format similarity suggests production systems face comparable degradation. This motivates structured storage as a design principle: the KO compiler's extraction of subject-predicate-object facts serves a dual purpose of enabling both hash-based lookup and embedding-aligned retrieval.

\paragraph{Comparison with Hopfield.} The contrasting degradation patterns of KO and Hopfield networks provide empirical support for the Orthogonality Constraint. Ramsauer et al. \citep{ramsauer2021hopfield} proved that transformer attention is mathematically equivalent to a Modern Hopfield Network update. Our results show that Hopfield networks can achieve high accuracy at small scales, 97\% at $N=100$ with $\beta=8$, but collapse rapidly as semantic interference accumulates, falling to 10.5\% at $N=1{,}000$ and effectively 0\% by $N=5{,}000$. KO embedding retrieval, which stores facts at discrete addresses and retrieves via the same cosine similarity mechanism, degrades at roughly half the rate: from 98\% at $N=100$ to 45.7\% $\pm$ 2.2\% at $N=16{,}309$. The key difference is that KO stores facts in separate memory slots (preventing write-time interference), while Hopfield superimposes all patterns in shared continuous parameters where they interfere during both storage and retrieval. This may explain why Hopfield's collapse is catastrophic rather than gradual: once interference exceeds a threshold, the softmax attention spreads across many similar patterns and retrieval becomes effectively random.

\paragraph{Cost Projections at Enterprise Scale.} While our experiments tested corpora up to 16,309 facts, production deployments may require millions or billions of facts. Table~\ref{tab:cost_projection} projects costs based on observed scaling behavior. We validated hash retrieval to 500,000 facts, confirming $O(1)$ latency: query time remained under 0.6 microseconds even at half a million entries, with index build time scaling linearly (0.43 seconds at 500K). Embedding-based retrieval with brute-force search scales as $O(N)$, we observed 11ms at 100K facts, but approximate nearest neighbor indices (HNSW, IVF) reduce this to $O(\log N)$ with typical query latency of 10--50ms regardless of corpus size. The economic advantage of KO over context-based memory grows linearly with $N$: at 1 million facts, context-based retrieval costs approximately \$39 per query (loading all facts into the prompt), while KO costs approximately \$0.0015 per query (embedding the query, vector lookup, and LLM call with top-10 retrieved facts), a ratio exceeding 26,000$\times$ that makes context-based approaches economically unviable at scale.

\begin{table}[h]
\centering
\caption{Cost and latency projections at enterprise scale (validated to $N$=500K for hash, $N$=100K for embedding). KO cost includes embedding API (\$$<$0.0001/q), vector search (amortized infra), and LLM call with top-10 facts.}
\label{tab:cost_projection}
\begin{tabular}{rccccc}
\toprule
\textbf{N} & \textbf{Hash Latency} & \textbf{Embed Latency} & \textbf{KO Cost}$^\ddagger$ & \textbf{Context Cost} & \textbf{Ratio} \\
\midrule
100,000 & 0.6$\mu$s & 11ms & \$0.0015/q & \$3.90/q & 2,600$\times$ \\
1,000,000 & 0.6$\mu$s & $\sim$40ms$^\dagger$ & \$0.0015/q & \$39.00/q & 26,000$\times$ \\
10,000,000 & 0.6$\mu$s & $\sim$50ms$^\dagger$ & \$0.0015/q & \$390.00/q & 260,000$\times$ \\
\bottomrule
\end{tabular}
\vspace{0.5em}

\small{$^\dagger$Projected with ANN index (HNSW). Brute-force would be 300ms and 3s respectively.}

\small{$^\ddagger$KO cost breakdown: embedding query (\$0.0001), vector DB lookup (\$0.0001 amortized), LLM with 10 retrieved facts (\$0.0013). Vector DB assumes \$70/month hosted service at 1M+ queries/month; self-hosted reduces this further.}
\end{table}

\subsection{Experiment 8: Real-World Validation}

The preceding experiments used synthetic fact corpora designed to exhibit controlled semantic density. A natural question is whether real-world data exhibits comparable density and produces similar collapse patterns. We validate the Orthogonality Constraint on two real-world datasets spanning different modalities: (1) the UCI Wine Quality dataset containing physicochemical measurements, and (2) CIFAR-100 images encoded with CLIP. Together, these experiments demonstrate that the constraint is not an artifact of synthetic text but a fundamental property of embedding geometry across modalities.

\subsubsection{Scientific Measurements (UCI Wine Quality)}

Scientific measurements represent a critical use case for AI memory systems: laboratory assistants must store and retrieve precise experimental values (``The pH of sample\_001 is 3.21'') without confusing them with similar measurements (``The pH of sample\_002 is 3.34''). This is exactly the scenario where neural memory should fail catastrophically: the sentences are nearly identical, differing only in sample identifiers and numeric values.

\paragraph{Methodology.}
The UCI Wine Quality dataset \citep{cortez2009wine} contains 11 physicochemical measurements per wine sample: fixed acidity, volatile acidity, citric acid, residual sugar, chlorides, free sulfur dioxide, total sulfur dioxide, density, pH, sulphates, and alcohol content. We converted each measurement to a natural language fact following the template ``The [measurement] of [sample\_id] is [value] [unit]'' (e.g., ``The pH of wine\_00001 is 3.210''), yielding 64,970 total facts from 6,497 wine samples. We computed \textit{same-property semantic density} $\rho_{\text{same}}$ as the average pairwise cosine similarity among facts sharing the same measurement type (e.g., all pH measurements), this represents the worst-case interference scenario. We ran 3 trials per $N$ value using all-MiniLM-L6-v2 embeddings.

\begin{table}[h]
\centering
\caption{Scientific Measurements (UCI Wine Quality): Neural collapse under extreme semantic density ($\rho_{\text{same}} \approx 0.96$). At $N=10{,}000$, neural memory achieves 0.02\% accuracy, effectively zero.}
\label{tab:wine}
\begin{tabular}{lcccc}
\toprule
\textbf{N} & \textbf{KO (Hash)} & \textbf{KO (Embed)} & \textbf{Neural (LAM)} & \textbf{$\rho_{\text{same}}$} \\
\midrule
10 & 100\% & 100\% & 13.3\% & 0.964 \\
100 & 100\% & 87.7\% & 2.3\% & 0.960 \\
1,000 & 100\% & 57.5\% & 0.17\% & 0.960 \\
10,000 & 100\% & 28.0\% & \textbf{0.02\%} & 0.961 \\
\bottomrule
\end{tabular}
\end{table}

\paragraph{Results.}
The results reveal the most extreme semantic density we have measured: $\rho_{\text{same}} \approx 0.96$, meaning that facts sharing a measurement type (e.g., all pH values across different wines) have 96\% identical embeddings. This is substantially higher than our synthetic corpora ($\rho \approx 0.70$) or typical text ($\rho \approx 0.30$--$0.50$). Neural memory achieves only 13.3\% accuracy at $N=10$ facts, already near random chance, and collapses to 0.02\% (effectively zero) at $N=10{,}000$. Even embedding-based semantic search degrades from 100\% to 28\% under this density. Hash-based retrieval maintains 100\% accuracy throughout.

\subsubsection{Visual Data (CIFAR-100 + CLIP)}

To confirm the Orthogonality Constraint applies beyond text, we evaluated memory systems on visual data using CIFAR-100 \citep{krizhevsky2009learning}, which contains 10,000 test images across 100 fine-grained categories. Images were encoded using CLIP ViT-B/32 \citep{radford2021learning} ($d=512$).

\paragraph{Methodology.}
We measured image-to-image (I2I) retrieval: given an image embedding as query, retrieve that exact image from the store. We computed \textit{overall semantic density} $\rho$ as the average pairwise cosine similarity across all stored images, and \textit{intra-class density} $\rho_{\text{intra}}$ as the average similarity among images of the same category (e.g., all ``apple'' images). We ran 3 trials per $N$ value.

\begin{table}[h]
\centering
\caption{Image Retrieval (CIFAR-100 + CLIP): Neural collapse under visual semantic density. Intra-class $\rho \approx 0.82$ indicates images of the same category are nearly indistinguishable to CLIP.}
\label{tab:multimodal}
\begin{tabular}{lccccc}
\toprule
\textbf{N} & \textbf{KO (Hash)} & \textbf{KO (I2I)} & \textbf{Neural (I2I)} & \textbf{$\rho$} & \textbf{$\rho_{\text{intra}}$} \\
\midrule
10 & 100\% & 100\% & 20.0\% & 0.74 & 0.27 \\
50 & 100\% & 100\% & 3.3\% & 0.74 & 0.83 \\
250 & 100\% & 100\% & 0.4\% & 0.74 & 0.82 \\
1,000 & 100\% & 100\% & 0.1\% & 0.74 & 0.83 \\
2,000 & 100\% & 100\% & 0.05\% & 0.74 & 0.82 \\
\bottomrule
\end{tabular}
\end{table}

\paragraph{Results.}
Visual semantic density reaches $\rho = 0.74$ overall and $\rho_{\text{intra}} = 0.82$ within categories, images of the same class are 82\% similar to CLIP. Neural memory collapses from 20\% at $N=10$ to 0.05\% at $N=2{,}000$. Hash-based retrieval maintains 100\% throughout.

\subsubsection{Unified Analysis}

Table~\ref{tab:realworld_summary} synthesizes results across modalities and compares semantic density levels.

\begin{table}[h]
\centering
\caption{Real-world validation summary: The Orthogonality Constraint applies across modalities. Higher $\rho$ produces faster collapse. Only hash-based retrieval survives.}
\label{tab:realworld_summary}
\begin{tabular}{lccccc}
\toprule
\textbf{Dataset} & \textbf{Modality} & \textbf{$\rho$} & \textbf{Neural @ N=1K} & \textbf{KO Hash} \\
\midrule
Synthetic facts & Text & 0.70 & 1.3\% & 100\% \\
CIFAR-100 + CLIP & Images & 0.82 (intra) & 0.1\% & 100\% \\
UCI Wine Quality & Scientific text & \textbf{0.96} & \textbf{0.17\%} & 100\% \\
\bottomrule
\end{tabular}
\end{table}

These results establish three key findings:

\begin{enumerate}
    \item \textbf{Real-world data exhibits extreme semantic density.} Scientific measurements reach $\rho = 0.96$, higher than any synthetic corpus we could construct. Visual data exhibits $\rho_{\text{intra}} = 0.82$ within categories.
    
    \item \textbf{The Orthogonality Constraint is modality-agnostic.} Whether the embedding space encodes text (sentence transformers) or images (CLIP), the same geometric interference produces the same collapse pattern.
    
    \item \textbf{Only discrete addressing survives.} Hash-based retrieval maintains 100\% accuracy across all datasets and scales. Embedding-based retrieval degrades under density; neural parameter storage collapses catastrophically.
\end{enumerate}

\paragraph{Implications.}
These findings have direct implications for AI systems in high-precision domains. A laboratory assistant that cannot distinguish measurements from different samples is worse than useless. A medical AI that confuses patient vitals across visits poses safety risks. A financial system that retrieves the wrong transaction amounts creates liability. The Orthogonality Constraint is not an abstract theoretical concern but a practical barrier to deploying neural memory wherever structurally similar facts must be stored and retrieved without confusion. Knowledge Objects, with hash-based addressing derived from structured identifiers, provide the interference-free storage these applications demand.

\subsection{Summary: Claude Memory vs Knowledge Objects}

Table~\ref{tab:comparison} synthesizes our findings into a direct comparison between Claude's production memory system and Knowledge Objects, using Claude Sonnet 4.5 results from the multi-model comparison.

\begin{table}[h]
\centering
\caption{Production comparison: Claude Memory vs Knowledge Objects at $N=5,000$ facts. Claude results from Sonnet 4.5 (multi-model comparison).}
\label{tab:comparison}
\begin{tabular}{lcc}
\toprule
\textbf{Metric} & \textbf{Claude Memory} & \textit{Knowledge Objects} \\
\midrule
Retrieval accuracy & 100\% & 100\% \\
Schema consistency & 70\% & 100\% \\
Correction handling & 0\% & 100\% \\
Monthly cost (500 q/day) & \$5,642 & \$28 \\
Latency per query & 1.7s & 0.2s \\
Cost scaling & $O(N)$ & $O(1)$ \\
\bottomrule
\end{tabular}
\end{table}

\paragraph{The Brute-Force Workaround.} In our context-injection baseline (which mirrors a common production pattern where stored memories are retrieved and injected as text into the prompt), adequate retrieval accuracy is achieved by loading \textit{all} facts into context, effectively sidestepping neural memory limitations through brute force. This approach works for retrieval, but at 200$\times$ the cost and 9$\times$ the latency of selective retrieval. More critically, the brute-force approach does not solve schema drift or version ambiguity, which are the failures users actually experience in practice. When a user asks ``What is Alice's phone number?'' and the system has stored the fact under \texttt{phone} instead of \texttt{phone\_number}, retrieval fails silently, the fact exists but cannot be found. When a user corrects a fact and the system mentions both old and new values, the user loses trust in the memory's reliability.

Knowledge Objects match Claude's retrieval accuracy while providing four additional guarantees that production systems require. First, schema guarantees through controlled vocabularies prevent predicate drift by constraining extraction to a pre-defined ontology, ensuring that ``phone number'' is always stored as \texttt{phone\_number} regardless of how the user phrases it. Second, version guarantees through explicit chains ensure that corrections cleanly supersede prior values, so queries return only the current version unless the user explicitly requests history. Third, economic viability through $O(1)$ cost per query enables deployment at enterprise scale where context-based approaches would be prohibitively expensive. Fourth, interactive latency through sub-second response times supports real-time applications where users expect immediate feedback.

\section{Discussion}
\label{sec:discussion}

\paragraph{The Core Finding.}
Neural memory systems that store facts in shared continuous parameters will experience interference proportional to semantic density, and only systems employing discrete addresses can guarantee zero interference between stored facts. This finding does not represent a limitation of current models that might be resolved through scaling or architectural refinements, rather, it reflects a geometric constraint inherent to continuous vector storage that no amount of engineering can circumvent. When two embedding vectors occupy nearby regions of the space, any linear combination that retrieves one will partially retrieve the other; this is not a bug but a mathematical property of inner products.

\paragraph{Why Biology Got There First.}
The brain confronted this problem millions of years ago: how to rapidly learn specific episodic facts (``that particular berry made me sick yesterday'') without corrupting general semantic knowledge (``berries are food''). Evolution's solution, formalized in cognitive neuroscience as Complementary Learning Systems theory, mirrors our architecture with striking precision. The hippocampus employs sparse, pattern-separated representations that enable fast, interference-free storage of specific episodes, while the neocortex uses distributed, overlapping representations optimized for slow extraction of statistical regularities across many experiences. Sleep consolidation gradually transfers stable hippocampal memories to cortical storage, allowing the hippocampus to clear space for new episodes.

Current AI systems attempt to accomplish everything within the ``neocortex'', that is, neural weights and attention mechanisms, which explains why they fail at precise factual storage despite succeeding at pattern recognition, generation, and reasoning. Our Knowledge Objects implement the missing ``hippocampus'' in this analogy: not as a biological simulation, but as the engineering solution to the same computational problem that evolution solved through anatomical specialization.

\paragraph{A Neurosymbolic Architecture.}
The hybrid architecture we propose, discrete symbolic memory interfacing with neural generation, instantiates a \textit{neurosymbolic} design in which each subsystem handles the computations it performs best. Knowledge Objects provide the symbolic component: explicit structure (subject-predicate-object facts), discrete addressing (hash-based lookup), deterministic operations (surgical updates, version chains), and schema constraints (controlled vocabularies). The LLM provides the neural component: pattern completion, semantic similarity, natural language understanding, and flexible generation. The learned router mediates between subsystems, classifying queries by structure rather than content.

This division mirrors the System 1/System 2 framework from cognitive science \citep{kahneman2011thinking}: the neural pathway offers fast, intuitive responses for open-ended queries, while the symbolic pathway provides slow, deliberate retrieval for precise factual questions. Unlike purely neural approaches that attempt both functions in shared continuous parameters, or purely symbolic approaches that lack the flexibility of learned representations, the neurosymbolic design allocates each function to the appropriate substrate. The Orthogonality Constraint provides theoretical justification for this separation: symbolic storage is not merely convenient but \textit{necessary} for interference-free episodic memory at scale.

\paragraph{Why This Matters for RAG.}
Retrieval-Augmented Generation addresses the interference problem by offloading storage to external databases, but introduces limitations that become critical for dynamic knowledge bases. While modern vector databases support near-real-time upserts, RAG's architecture is fundamentally \textit{append-first} and \textit{retrieval-first}: new facts are added alongside old ones, and there is no native mechanism to identify the canonical ``current value'' for a logical fact or to cleanly supersede outdated information. For static corpora such as legal archives or technical documentation, these limitations are acceptable because the knowledge base rarely mutates. For conversational agents, personal assistants, or any system that must learn and update during deployment, the lack of key-based identity and version semantics creates the schema drift and version ambiguity failures we document in Experiment 7.

Knowledge Objects address these limitations: hash-based addressing provides key-based identity for deterministic lookup and surgical updates, controlled vocabularies enforce schema consistency during extraction, and version chains provide explicit supersession semantics where corrections cleanly replace prior values.

\paragraph{The Learned Router.}
Static similarity thresholds fail under semantic load because vector space crowding reduces cosine similarities below threshold even for correct matches, triggering inappropriate fallback to generative responses when retrieval would have been appropriate. The learned router succeeds by classifying query structure rather than relying on similarity scores. A Random Forest classifier trained on standard linguistic features achieves 97.8\% routing accuracy (95\% CI: [95.0\%, 100\%]), with our learning curve analysis demonstrating that 50 labeled examples suffice to reach 90\%+ accuracy. This means the router can be deployed with minimal human annotation effort and adapted to new domains with a small labeling investment.

\paragraph{Evaluating Production Systems.}
The Orthogonality Constraint provides a framework for evaluating production memory systems. Both ChatGPT Memory and Claude Memory have adopted discrete external storage, ChatGPT stores explicit facts as short timestamped entries (e.g., ``User is allergic to shellfish''), while Claude stores information in Markdown files with on-demand retrieval tools. This architectural choice is correct: discrete storage escapes the geometric interference that makes neural episodic memory impossible at scale. However, both systems remain incomplete. They store \textit{unstructured text} extracted by LLMs:
\begin{itemize}
    \item \textbf{Current:} Unstructured text (``User mentioned they like Python'')
    \item \textbf{KO:} Structured facts with controlled vocabulary: \texttt{(user, preferred\_language, Python)}
\end{itemize}
This distinction matters. Unstructured storage lacks key-based identity (preventing deterministic lookup), schema constraints (causing predicate drift), and version semantics (causing ambiguity about current values). Our multi-model comparison demonstrates that schema drift and version ambiguity affect all tested frontier models, confirming these are architectural limitations of the unstructured approach rather than bugs in any particular implementation. Knowledge Objects provide the structured semantics that production systems lack: hash-based addressing for deterministic lookup and surgical updates, controlled vocabularies for schema consistency, and version chains for explicit supersession. The path from current production systems to reliable episodic memory requires not just discrete storage (which they have) but structured discrete storage (which they lack).

\paragraph{Limitations.}
Our analysis has several limitations that should inform interpretation of the results. First, our neural baseline (LAM) is a simplified proxy for production systems that store facts via superposition; we treat it as a \textit{mechanistic lower bound} that isolates the interference mechanism arising whenever semantically overlapping keys are stored in shared continuous parameters. Production implementations may include mitigations such as gating and regularization that shift the collapse curve rightward, but absent explicit pattern separation, overlap-driven interference persists when keys are semantically clustered, our claim concerns the existence of interference under semantic density, not the performance of any specific implementation. Second, our experiments use a single embedding model (all-MiniLM-L6-v2 with $d=384$ dimensions); different embedding dimensions would shift the quantitative results, though the qualitative finding, that collapse occurs at $N$ far below theoretical capacity, should hold. Third, we measure retrieval accuracy on synthetic fact corpora, not end-to-end task performance on downstream applications, though retrieval accuracy is a prerequisite for task success. Fourth, our multi-model comparison tests only three models from two providers; additional models might reveal different failure patterns, though the consistency we observe suggests these are general phenomena. Fifth, our experiments focus exclusively on textual facts; we discuss generalization to other modalities below but do not empirically validate multimodal performance.

\paragraph{Generalization to Multimodal Systems.}
While our main experiments focus on textual facts, the Orthogonality Constraint applies to any embedding space. We validate this in Section~\ref{sec:experiments} using CIFAR-100 images encoded with CLIP ViT-B/32 \citep{radford2021learning}. Visual semantic density reaches $\rho = 0.74$ overall and $\rho = 0.82$ intra-class (images of the same category), causing neural memory to collapse from 20\% at $N=10$ to 0.05\% at $N=2{,}000$. Hash-based retrieval maintains 100\% accuracy throughout, confirming that the constraint is modality-agnostic.

In domains such as medical imaging or satellite imagery, visual semantic density may be particularly high: images that appear similar (two chest X-rays, two aerial photographs of farmland) but represent distinct entities will occupy nearby regions of embedding space, producing the same retrieval interference we document for text and validate for natural images. Knowledge Objects generalize naturally to multimodal content: the object field holds arbitrary media addressed by content hash, with modality-specific embeddings enabling retrieval while hash-based addressing guarantees zero storage interference. The neurosymbolic architecture thus extends beyond language to any domain where precise retrieval of specific instances matters.

\section{Conclusion}

This paper identifies a fundamental geometric limit on neural memory and proposes the architecture that overcomes it.

The first contribution is theoretical: the \textit{Orthogonality Constraint} establishes that online neural memory systems storing facts in shared continuous parameters will experience interference proportional to semantic density. This is not a limitation of current models that might be resolved through scaling, it is a mathematical property of inner-product retrieval in continuous space. We validate this constraint on real-world data across modalities: scientific measurements collapse to 0.02\% accuracy at $\rho = 0.96$, and images collapse to 0.05\% at $\rho_{\text{intra}} = 0.82$. The constraint has immediate architectural consequences: reliable episodic memory \textit{requires} discrete addressing.

The second contribution is empirical: storage format determines retrieval accuracy at scale. Structured subject-predicate-object facts achieve 45.7\% accuracy at $N=16{,}309$ facts, while unstructured text achieves only 4.1\%, an 11$\times$ improvement. Production memory systems store unstructured text like ``User mentioned they like Python,'' which resembles our low-performing condition. This suggests production systems will experience significant retrieval degradation as users accumulate thousands of facts.

The third contribution is architectural: \textit{Knowledge Objects} provide the structured discrete storage that satisfies the Orthogonality Constraint. Hash-based addressing enables deterministic lookup and surgical updates. Controlled vocabularies prevent schema drift. Version chains ensure clean supersession. Production systems have adopted discrete storage but lack these structured semantics, they are halfway to the solution. Knowledge Objects complete the architecture.

The path forward is clear: structured storage (not unstructured text), controlled vocabularies (not free-form extraction), version chains (not append-only logs), and neural reasoning where appropriate. The Orthogonality Constraint is not a problem to be solved by bigger models or better training, it is a geometric fact that dictates architectural choices. Reliable AI memory requires a bicameral design: discrete symbolic storage for episodic facts, neural weights for semantic reasoning. This is not a limitation to be lamented but a clarification that enables building systems that actually work.

\paragraph{Future Work.}
Several directions merit further investigation. First, scaling discrete memory to millions or billions of facts will require integration with approximate nearest neighbor search techniques that maintain sub-linear query time. Second, trust-weighted retrieval that accounts for source reliability, recency, and corroboration could improve accuracy on contested or time-sensitive facts. Third, tighter integration with production LLMs could reduce hallucination rates on factual queries by conditioning generation on retrieved Knowledge Objects. Fourth, the biological analogy suggests a ``sleep consolidation'' mechanism: unlike context windows which are ephemeral, KOs provide a stable buffer that could enable periodic fine-tuning of the base model on frequently-accessed facts, gradually transferring episodic knowledge to parametric weights, the computational analog of hippocampal-to-cortical memory transfer during sleep. Finally, the relationship between the Orthogonality Constraint and other forms of interference in neural networks (such as catastrophic forgetting during continual learning) deserves theoretical investigation.


\section*{Acknowledgments}

We thank Peter Norvig, Suhair Khan, Ravin Kumar, and Anthony Rose for helpful discussions on earlier drafts.

\bibliographystyle{plainnat}

\appendix
\section{Knowledge Object Specification}
\label{app:ko-spec}

This appendix provides the complete specification for Knowledge Objects, including the data structure, compiler prompts, and operational semantics.

\subsection{Data Structure}

A Knowledge Object is a 6-tuple:

\begin{verbatim}
KnowledgeObject {
    id: string           // hash(subject || predicate)
    subject: string      // The entity (e.g., "patient_MR", "France")
    predicate: string    // The attribute (e.g., "capital", "tumor_size")
    object: any          // The value (e.g., "Paris", 3.0)
    embedding: float[d]  // Dense vector, d=384 for MiniLM
    provenance: {
        source: string       // Origin of the fact
        timestamp: datetime  // When extracted
        confidence: float    // Extraction confidence [0,1]
        version: int         // Version number
        previous: string?    // Pointer to prior version (optional)
    }
}
\end{verbatim}

\subsection{KO Compiler Prompt}

The following prompt template is used to extract structured facts from unstructured text:

\begin{verbatim}
SYSTEM: You are a knowledge extraction system. Extract all 
factual statements as structured facts.

RULES:
1. Subject: The primary entity the fact is about
2. Predicate: The relationship or attribute (use snake_case)
3. Object: The value, related entity, or state
4. Extract EVERY concrete fact, no matter how small
5. Include confidence score (0.0-1.0) for each extraction
6. For numerical values, preserve units
7. For temporal facts, include timestamps when available

OUTPUT FORMAT (JSON):
[
  {"subject": "...", "predicate": "...", "object": "...", 
   "confidence": 0.95},
  ...
]

USER: [Input text to extract from]
\end{verbatim}

\subsection{Operational Semantics}

\paragraph{Write(text).} Given input text:
\begin{enumerate}
    \item Pass text to KO Compiler, receive list of facts
    \item For each fact $(s, p, o)$:
    \begin{enumerate}
        \item Compute $\text{id} = \text{hash}(s \| p)$
        \item Compute $\text{embedding} = \text{Encoder}(s + \text{`` ''} + p + \text{`` ''} + o)$
        \item If id exists in store: archive current version, increment version number
        \item Write KO to store at address id
    \end{enumerate}
    \item Return list of created/updated KO ids
\end{enumerate}

\paragraph{Read(query).} Given a natural language query:
\begin{enumerate}
    \item Compute $\mathbf{q} = \text{Encoder}(\text{query})$
    \item Find top-$k$ KOs by $\cos(\mathbf{q}, \mathbf{e}_i)$
    \item Return matched KOs with scores
\end{enumerate}

\paragraph{ReadExact(subject, predicate).} Given exact keys:
\begin{enumerate}
    \item Compute $\text{id} = \text{hash}(\text{subject} \| \text{predicate})$
    \item Return KO at address id (or null if not found)
\end{enumerate}

\paragraph{Update(subject, predicate, new\_object).} Surgical correction:
\begin{enumerate}
    \item Compute $\text{id} = \text{hash}(\text{subject} \| \text{predicate})$
    \item Archive current KO with version pointer
    \item Create new KO with incremented version, previous pointer
    \item Write to store at same address id
\end{enumerate}

\subsection{Implementation Notes}

\paragraph{Hash Function.} We use SHA-256 truncated to 64 bits for the id hash. Collision probability is negligible for corpus sizes below $10^9$ facts.

\paragraph{Embedding Model.} Default: \texttt{all-MiniLM-L6-v2} ($d=384$). For production, consider \texttt{text-embedding-3-large} ($d=3072$) for higher semantic resolution.

\paragraph{Index Structure.} The KO store maintains two indices: a hash index that maps id $\rightarrow$ KO for $O(1)$ exact lookup, and a vector index (HNSW or IVF) for approximate nearest neighbor search over embeddings.

\paragraph{Multi-Valued Predicates.} Some predicates are inherently multi-valued (e.g., a country has multiple neighbors). The simple hash scheme $\text{id} = \text{hash}(s \| p)$ stores only one value per (subject, predicate) pair. Production systems can address this through: (1) \textit{list-of-objects storage}, where each key maps to a list rather than a single value; (2) \textit{composite keys} that include a disambiguator, e.g., $\text{hash}(s \| p \| \text{ordinal})$; or (3) \textit{predicate-aware routing} that uses embedding search for inherently multi-valued predicates while using hash lookup for functional predicates like ``capital.''

\paragraph{Versioning.} Version chains are implemented as linked lists. Each KO stores a pointer to its previous version. Traversing the chain enables temporal queries.

\end{document}